\documentclass[acmsmall,review=false,anonymous=false]{acmart}
\pdfoutput=1 

\usepackage{std}

\setcopyright{none}

\begin{document}

\title{A Language-based Serverless Function Accelerator}

\author{Emily Herbert}
\email{emilyherbert@cs.umass.edu}
\affiliation{%
  \institution{University of Massachusetts Amherst}
}

\author{Arjun Guha}
\email{arjun@cs.umass.edu}
\affiliation{%
  \institution{University of Massachusetts Amherst}
}

\begin{abstract}

\emph{Serverless computing} is an approach to cloud computing that allows
programmers to run \emph{serverless functions} in response to external events.
Serverless functions are priced at sub-second granularity, support transparent
elasticity, and relieve programmers from managing the operating system. Thus
serverless functions allow programmers to focus on writing application code,
and the cloud provider to manage computing resources globally. Unfortunately,
today's serverless platforms exhibit high latency, because it is difficult to
maximize resource utilization while minimizing operating costs.

This paper presents \emph{serverless function acceleration}, which is an
approach that transparently lowers the latency and resource utilization of a
large class of serverless functions. We accomplish this using language-based
sandboxing, whereas existing serverless platforms employ more expensive
operating system sandboxing technologies, such as containers and virtual
machines. OS-based sandboxing techniques are compatible with more programs than
language-based techniques. However, instead of ruling out any programs, we use
language-based sandboxing when possible, and operating system sandboxing if
necessary. Moreover, we seamlessly transition between language-based and
OS-based sandboxing by leveraging the fact that serverless functions must
tolerate re-execution for fault tolerance. Therefore, when a serverless function
attempts to perform an unsupported operation in the language-based sandbox, we
can safely re-execute it in a container.

Security is critical in cloud computing, thus we present a serverless function
accelerator with a minimal trusted computing base (TCB). We use a new approach
to trace compilation to build a source-level, interprocedural, execution trace
tree for serverless functions written in JavaScript. We compile trace trees to
a safe subset of Rust, validate the compiler output, and link it to a runtime
system. The tracing system and compiler are untrusted, whereas the trusted
runtime system and validator are less than 3,200 LOC of Rust.

We evaluate these techniques in our implementation, which we call \sysname{},
and show that our approach can significantly decrease the latency and resource
utilization of serverless functions, e.g., increasing throughput of I/O bound
functions by \dataHeadlineThroughput. We also show that the impact of tracing
is negligible and that \sysname{} can seamlessly switch between its two modes
of sandboxing.
\end{abstract}

\maketitle

\section{Introduction}
\label{introduction}

\emph{Serverless computing} is a recent approach to cloud-computing that allows
programmers to run small, short-lived programs, known as \emph{serverless
functions}, in response to external events. In contrast to rented virtual
machines, serverless computing is priced at sub-second granularity and the
programmer only incurs costs when a function is processing an event. The
serverless platform fully manages the (virtualized) operating system,
load-balancing, and auto-scaling for the programmer. In particular, the
platform transparently starts and stops concurrent instances of a serverless
function as demand rises and falls. Moreover, the platform terminates all
instances of a function if it does not receive events for an extended period of
time.

Unfortunately, today's serverless platforms exhibit high tail
latency~\cite{shahrad:micro-faas}. This problem occurs because the serverless
platform has to make a tradeoff between maximizing resource utilization (to
lower costs) and minimizing event-processing latency (which requires idle
resources). Therefore, an approach that simultaneously lowers latency and
resource utilization would have several positive effects, including lowering
cold start times and lowering the cost of keeping idle functions resident.

\paragraph{The dynamic language bottleneck} A key performance bottleneck for
serverless functions is that they are typically written in dynamic languages,
such as JavaScript. Contemporary JavaScript virtual machines are
state-of-the-art JITs that make JavaScript run significantly faster than
simpler bytecode interpreters~\cite{deutsch:efficient-smalltalk}. Nevertheless,
JIT-based virtual machines are not ideal for serverless computing for several
reasons. First, their instrumentation, optimization, and dynamically generated
code can consume a significant amount of time and
memory~\cite{dean:selective-specialization}. Second, a JIT takes time to reach
peak performance, and may never reach peak performance at
all~\cite{barrett:vm-warmup}. Finally, existing language runtimes require an
operating system sandbox. In particular, Node---the de facto standard for
running JavaScript outside the browser---is not a reliable
sandbox~\cite{brown:js-binding-bugs}.

\paragraph{Alternative languages}

An alternative approach is to give up on JavaScript, and instead require the
programmer to use a language that performs better and is easier to secure in a
serverless execution environment. For example, Boucher et al. present a
platform that only supports serverless functions written in
Rust~\cite{microin}. This allows them to leverage Rust's
language-level guarantees to run several serverless functions in a
single shared process, which is more lightweight than per-process sandboxing or
per-container sandboxing. However, Rust is not a panacea. For many programmers,
Rust has a steep learning curve, and a deeper problem
is that Rust's notion of safety is not strong enough for serverless computing.
Even if a program is restricted to a safe subset of Rust, the language
\emph{does not} guarantee resource isolation, deadlock freedom, memory leak
freedom, and other critical safety properties~\cite{rust-unsafe}. Boucher et
al. identify these problems, but are not able to address them in full.

\paragraph{Compiling JavaScript Ahead-of-Time}

Consider a small variation of the previous idea: the serverless platform could
compile JavaScript to Rust for serverless execution. JavaScript would make the
platform appeal to a broader audience, the Rust language would ensure
memory-safety, and the JS-to-Rust compiler could insert dynamic checks to
provide guarantees that Rust does not statically provide. Unfortunately, this
approach would run into several problems. First, Garbage-collected languages
support programming patterns that cannot be expressed without a garbage
collector~\cite[p. 9]{jones:gc}. Therefore, many JavaScript programs could not
be compiled without implementing garbage collection in Rust, which requires
unsafe code (i.e., to traverse stack roots). Second, dynamically typed
languages support programming patterns that statically typed languages do
not~\cite{thf:typedscheme,guha:flowtypes,furr:drubyoopsla,chugh:djs}.
Therefore, a JS-to-Rust compiler would have to produce Rust code that is
littered with type-checks and type-conversions~\cite{scheme-to-ml}, which would
be slower than a JIT that eliminates type-checks based on runtime type
feedback~\cite{holtz:type-feedback}. Finally, JavaScript has several obscure
language features (e.g., proxy objects and the lack of
arity-checks)~\cite{guha:js,bodin:coqjs,maffeis:jssemantics} that are difficult
to optimize ahead-of-time. Although recent research has narrowed the gap
between JIT and AOT compilers~\cite{serrano:js-aot}, JITs remain the fastest
way to run JavaScript.

\paragraph{Our approach}

The aforementioned approaches assume serverless functions are arbitrary programs,
and overlook some unique properties that we can exploit:

\begin{enumerate}

\item A typical serverless function is \emph{short lived} and \emph{consumes
limited memory}. For example, a study of serverless workloads on Azure found
that 50\% of all serverless functions process events in less than one second
(on average), and consume less than 170 MB of
memory~\cite{shahrad:serverless-in-the-wild}. This is to be expected, because
serverless functions often respond to events triggered by end-users of
interactive systems.

\item A serverless function has \emph{transient in-memory state}, and must place
persistent state in external storage. This allows the function to tolerate faults
in the (distributed) serverless execution platform, and allows the platform
to evict a running function an any time without notification.

\item A serverless function is \emph{idempotent}, which means it must tolerate
re-execution, e.g., using transaction IDs to avoid duplicating side-effects.
This allows the serverless platform to naively re-invoke a function when 
it detects a potential fault.

\end{enumerate}

This paper presents \sysname{}, which is a \emph{serverless function
accelerator}, which uses a language-based sandbox to accelerate serverless
functions written in JavaScript, instead of operating system sandboxing, which
is used today. Ordinarily, moving to a language-based sandbox would restrict
what programs can do. For example, today's serverless functions can embed shell
scripts, launch binary executables, write to the filesystem, and so on, within
the confines of an operation-system sandbox (e.g., a Docker container).

However, instead of asking the programmer to chose between the two sandboxing
modes, \sysname{} uses language-based sandboxing when possible, and
\emph{transparently} falls back to container-based sandboxing if necessary.
This approach works because serverless functions must be \emph{idempotent}.
Apart from the difference in performance, a programmer cannot write code that
observes if the function is running in our new language-based sandbox or the
usual container-based sandbox. For example, suppose a function running in the
language-based sandbox attempts to run a shell script. In this case, \sysname{}
terminates the language-based sandbox, and re-executes the function in a
container with a virtual filesystem. The programmer will observe high latency
for that request, which could be caused by a number of factors. Moreover, the
\sysname{} runtime will determine that future executions of the function should
use container-based sandboxing to avoid needless re-execution.

\sysname{} also eschews garbage collection, and instead uses an an arena
allocator that frees memory after each response. This approach
is safe, because serverless functions must tolerate \emph{transient in-memory
state}.

Security is another factor that affects the design of \sysname{}. \sysname{} is
built in Rust and is carefully designed to minimize the trusted computing base
(TCB). For language-based sandboxing, \sysname{} generates Rust code from
JavaScript. This shifts a significant portion of the TCB out of our
implementation and onto the Rust type system, which has been heavily studied
using formal methods~\cite{jung:rustbelt}. However, \sysname{} is \emph{not} a
general-purpose JS-to-Rust compiler. As discussed above, a JS-to-Rust compiler would
suffer several pitfalls due to the ``impedance mismatch'' between the two
languages (e.g., types and garbage collection). Instead, \sysname{} first
instruments the source code of a serverless function to dynamically generate
 an \emph{inter-procedural execution trace tree}, which we
compile to Rust. This approach is closely related to tracing JIT compilers.
However, a unique feature of our trace tree is that it includes asynchronous callbacks. To the
best of our knowledge, all prior JITs are limited to sequential code. However,
the ``hot path'' in a typical serverless function includes asynchronous web
requests, thus we have to develop this capability.

Tracing in \sysname{} thus works as follows. The serverless function begins
execution in a container, with its source code instrumented to dynamically
build an execution trace tree. After a number of events, \sysname{} extracts
the trace tree and compiles it to Rust. Subsequent events are thus processed
more efficiently in Rust instead of the container. If the Rust code receives an
event that triggers an unknown execution path, it aborts and falls back to the
container. However, whereas a general-purpose JIT must use sophisticated
techniques such as deoptimization and on-stack replacement, \sysname{} can
naively abort the fast-path (Rust) and re-execute the program in the slow-path
(container).

We evaluate \sysname{} with a suite of \dataNumBenchmarks{} typical serverless
functions and show that \sysname{} 1)~reduces resource usage, which
allows it to handle more concurrent requests; 2)~reduces the latency of
serverless functions; and 3)~seamlessly transitions between its two sandboxing
modes.

\paragraph{Contributions} To summarize we make the following contributions.

\begin{enumerate}

  \item We show that it is possible to transparently accelerate serverless
  functions using language-based techniques, by exploiting the fact that
  serverless functions are idempotent and have transient in-memory
  state.

  \item We present a source-to-source compiler and runtime system that
  instruments JavaScript code, to dynamically generate an inter-procedural
  execution trace tree. A unique feature of our approach to tracing is that it
  includes asynchronous callbacks. In addition, our approach to source-level
  tracing uses a runtime system that grows the trace using zipper-like
  operations~\cite{huet:zipper}.

  \item We present a compiler that translates trace trees to a safe subset of
  Rust, which minimizes the amount of new code that the serverless platform has
  to trust.

  \item We evaluate \sysname{} on \dataNumBenchmarks{} canonical serverless
  functions. We show that it can increase the throughput of
  serverless functions by \dataHeadlineThroughput{}, can reduce CPU utilization
  by a factor of \dataHeadlineCPU{}, and may help alleviate the cold start
  problem.

\end{enumerate}

\ifsubmission
We will make our code and data publicly available under an open-source license,
and would submit \sysname{} for artifact evaluation.
\fi

The rest of this paper is organized as follows. \Cref{intro-to-serverless}
introduces serverless computing and the \sysname{} API. \Cref{tracing} presents
the language of trace trees and describes how we construct traces from JavaScript.
\Cref{traces-to-rust} presents the trace-to-Rust compiler. \Cref{rts} presents the
\sysname{} invoker, which manages both containers and language-based sandboxes.
\Cref{evaluation} evaluates \sysname{}. \Cref{discussion} discusses the
security of the \sysname{} design. \Cref{related} discusses related work.
Finally, \Cref{conclusion} concludes.

\section{Serverless Programming with \sysname{}}
\label{intro-to-serverless}

In this section we introduce the serverless programming model, using the
\sysname{} API. We then discuss the design and implementation of
traditional, container-based serverless platforms, which is relevant
to the design of \sysname{}.

\subsection{Programming with \sysname{}}

\begin{figure}
\begin{minipage}{0.46\columnwidth}
\centering
\lstset{language=JavaScript}
\begin{lstlisting}
let c = require('containerless');

function main(req) {
  function F(resp) {
    let u = req.body.username;
    let p = req.body.password;
    if (resp[u] === p) {
      c.respond('ok');
    } else {
      c.respond('error');
    }
  }
  c.get('passwords.json', F);
}
\end{lstlisting}
\caption{A serverless function to authenticate users. The \sysname{} API is
similar to the APIs provided by commercial serverless computing platforms.}
\label{login-example-js}
\end{minipage}\hfill
\begin{minipage}{0.46\columnwidth}
\centering
\includegraphics[width=\columnwidth]{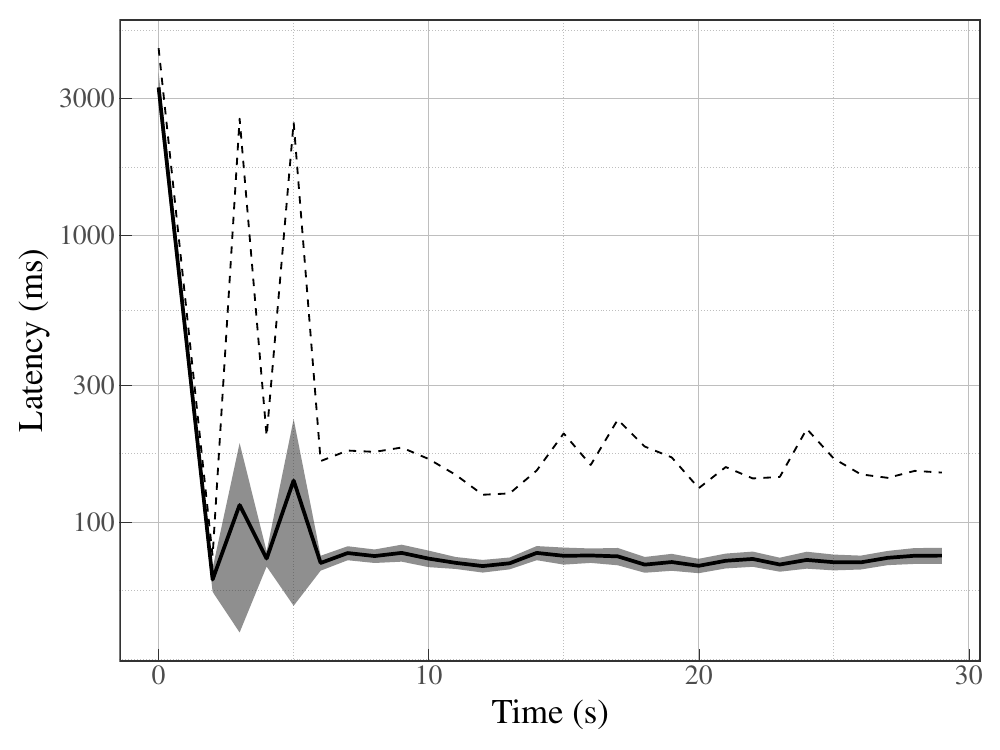}
\caption{Latency observed from a series of requests sent to a function hosted on
Google Cloud Platform. The solid lines show the mean response latency, with the
95\% confidence interval depicted by the shaded region around the mean. The
dotted lines show the maximum latency.}
\label{gcp-plot}
\end{minipage}

\end{figure}

\Cref{login-example-js} shows an example of a serverless function, written
with \sysname{}, that authenticates users. We note that `function'' is a 
misnomer, since a serverless function is in fact a serverless program, with helper 
functions, multiple modules, dependencies, etc. For consistency with the literature, 
we refer to serverless programs as serverless functions.

The code is written in JavaScript and uses the \sysname{} API.
The global \lstinline|main| function is the entrypoint, and it
receives a web request carrying a username and password (\lstinline|req|). The
function then fetches a dictionary of known users and their passwords from cloud
storage (\lstinline|resp|), validates the received username and password, and
then responds with \lstinline|'ok'| or \lstinline|'error'|.

The function illustrates an important detail: JavaScript does not support
blocking I/O. Therefore, all I/O operations take a callback function and return
immediately. For example, the \lstinline|c.get| function takes two arguments: a
URL to get, and a callback function that eventually receives the response.
Therefore, the \lstinline|main| function is also asynchronous. To return a
response, the serverless function calls \lstinline|c.respond| within a
callback. All JavaScript-based serverless programming platforms have similar
APIs that either use callback functions or promises.\footnote{We believe that
with some engineering effort, it should be possible to mimic the API of an
existing serverless platform (\cref{discussion}).}

The design of this serverless function is similar to a simple web server.
However, some key differences are that the function does not choose a listening
port or decode the request. The serverless platform manages these low-level
details for the programmer. In this case, when the programmer creates this
function, the platform assigns it a unique URL, and runs the function to
respond to requests at that URL. The platform also manages the operating system
and JavaScript runtime (including security updates), collects execution logs,
and provides other convenient features.

\subsection{Design and Implementation of Traditional Serverless Platforms}

A serverless platform involves several components running in a
distributed system. For example, OpenWhisk, which is the open-source serverless
platform underlying IBM Cloud Functions, relies on a web frontend, an
authentication database, a load balancer, and a message bus, all to process a
single event~\cite{shahrad:micro-faas}.

Our work focuses on the \emph{invoker}, which is the component that receives
events for serverless functions, and forwards them to a pool of containers that
it uses to execute serverless functions in isolation. The invoker places
resource limits (e.g., CPU and memory limits) on all containers, and runs one
function in each container. Within each container, the serverless function runs
in a process that receives and responds to events (usually over the container's
virtual network). For functions written in JavaScript, the process is a Node
process.

A single invoker can handle several concurrent events. Moreover, an event may
trigger one of several serverless functions from different customers, and the
set of functions may change over time. The invoker may have several
containers running concurrently for a single function, in which
case it manages load-balancing across the running containers. If an error
occurs during event processing (e.g., a container is not reachable on
the network), the invoker hides the fault and re-sends the event to another
container. 

A \emph{warm start} occurs when the invoker receives an event for a function
$f$, and it has an idle container with $f$'s code. In contrast, a \emph{cold
start} occurs when the invoker needs to create a new container, either because
the event triggers a function that has not recently run, or because all
existing containers for $f$ are busy. Cold starts incur significant overhead
compared to warm starts, and result in high tail latency. \Cref{gcp-plot} shows
the latency observed while sending series of requests to a function hosted on
Google Cloud Platform. The effects of cold start can be observed through an
initial 5 seconds after the first request.

Unfortunately, cold starts are unavoidable. It is not cost-effective for a cloud
platform to always have idle containers for every function.
Moreover, the invoker has to evict idle
containers after a period of time to allocate resources to other functions.
Futhermore, it is unsafe to reuse an idle container for a function $f$ to handle
an event for a function $g$: doing so risks leaking data from one customer to
another via operating system resources (e.g., temporary files). Finally,
the invoker cannot run two concurrent processes from two different customers in the
same container. Instead, the invoker ensures that a single container only ever
processes events for a single function.\footnote{SAND~\cite{istemi:sand}
proposes running multiple events in a single container, as long as they service
the same customer. Thus customers have to ensure that their functions do not
interfere with each other.}

In summary, the serverless platform automatically takes care of load-balancing,
failure recovery, and resource allocation for the programmer. Moreover, since
it uses a shared pool of computing resources, thus an individual programmer
does not have to pay for idle computing time.

However, the serverless abstraction is not transparent, and the programmer has
to ensure that their serverless function satisfies some key
properties~\cite{jangda:lambda-lambda,obetz:serverless-events}. 1)~When
the platform detects a failure while handling an event, it simply re-invokes a
container. For functions with external effects (e.g., writes to an external
database), it is up to the programmer to ensure that the function is
idempotent, so that re-execution is safe. 2)~When an invoker evicts an
idle function, it does so without any notification. Therefore, functions have
\emph{transient in-memory and on-disk state}. Programmers have to ensure that
all persistent state is saved to external storage (e.g., cloud storage or a
cloud-hosted database). 3)~To manage resources, the platform imposes a
\emph{hard timeout} on all functions (at most a few minutes on current
platforms). If a programmer needs to perform a lengthier computation, they need
to break it up into smaller functions. These are the characteristics that
\sysname{} exploits for serverless function acceleration.

\section{From JavaScript to Dynamic Trace Trees}
\label{tracing}

This section presents how \sysname{} turns a serverless function into a
dynamically generated trace tree. \Cref{traces-to-rust} describes the
trace-to-Rust compiler.

\begin{figure}
\footnotesize
\(
\begin{array}{@{}rcll@{}r@{}}
\multicolumn{4}{@{}l}{\textbf{Operators}} \\
\mathit{op} & ::= & \jsOp{+} \mid \jsOp{-} \mid \jsOp{*} \mid \cdots \\
\multicolumn{4}{@{}l}{\textbf{Expressions}} \\
\jsExpr{} & ::=  & c & \textrm{Constant} \\
          & \mid & x & \textrm{Variable} \\
          & \mid & \jsExpr{}_1~\mathit{op}~\jsExpr_2 & \textrm{Binary operation}  \\
\multicolumn{4}{@{}l}{\textbf{Binding Forms}} \\
\jsB & ::=  & \jsExpr  & \textrm{Expression} \\
     & \mid & \jsKw{function}\jsOp{(}x_1\cdots x_n\jsOp{)}~\jsBlk
            & \textrm{Abstraction} \\
     & \mid & f\jsOp{(}e_1\cdots e_n\jsOp{)}
            & \textrm{Application}
\end{array}
\)
\quad\vrule\quad
\(
\begin{array}{@{}rcll@{}r@{}}
\multicolumn{4}{@{}l}{\textbf{Block}} \\
\jsBlk & ::= & \jsOp{\{}~\jsStmt_1 \cdots \jsStmt_n\jsOp{\}} \\
\multicolumn{4}{@{}l}{\textbf{Statements}} \\
\jsStmt & ::=  & \jsKw{let}~x~\jsOp{=}~\jsB\jsOp{;} 
               & \textrm{Binding} \\
        & \mid & \jsBlk
               & \textrm{Block} \\
        & \mid & \jsKw{if}~\jsOp{(}\jsExpr\jsOp{)}~\jsStmt_1~\jsKw{else}~\jsStmt_2
               & \textrm{Conditional} \\
        & \mid & \jsKw{while}~\jsOp{(}\jsExpr\jsOp{)}~\jsStmt
                & \textrm{Loop} \\
        & \mid & x~\jsOp{=}~\jsB\jsOp{;}
                & \textrm{Assignment} \\
        & \mid & \jsLabel{\ell}\jsOp{:}~\jsStmt
                & \textrm{Label} \\
        & \mid & \jsKw{break}~\jsLabel{\ell}\jsOp{;}
                & \textrm{Break} \\
        & \mid & \jsKw{return}~\jsExpr\jsOp{;}
                & \textrm{Return} 
\end{array}
\)
  
\caption{The fragment of JavaScript that we use to present tracing.
\sysname{} supports many other JavaScript features.}

\label{js-anf-syntax}
\end{figure}

\subsection{A Representative Fragment of JavaScript}
\lstset{language=JavaScript}

\Cref{js-anf-syntax} presents a small fragment of JavaScript, which includes
first-class functions, assignable variables, conditionals, while loops, labels,
and breaks. We also restrict the syntax of JavaScript so that all functions
definitions and applications are named (similar to A Normal Form~\cite{anf}).
This fragment of JavaScript allows us to present the essentials of our approach
to trace generation in the rest of this section. \sysname{} generates traces
for the rest of JavaScript in two ways. 1)~The implementation natively supports
a variety of features including objects (with prototype inheritance), arrays,
and all JavaScript operators. These features do not affect the control-flow of
a program, thus trace generation is routine. 2)~\sysname{} supports many more
features by translating them into equivalent features. For example, we
translate \lstinline|for| to \lstinline|while|,
\lstinline|switch| to \lstinline|if|, generate fresh
names for anonymous functions, and so on.

\sysname{} does not support 1)~getters and setters 2)~\lstinline|eval|, and 3)
newer reflective and metaprogramming features such as object proxies. We
believe that it would be possible to getters, setters, and metaprogramming
features to work with more engineering effort. However,
\lstinline|eval|---since it allows dynamically loading new code---is the only
feature that is fundamentally at odds with our approach. If a program uses
\lstinline|eval|, we abort tracing and fall back to using containers.

\begin{figure}

\footnotesize

\(
\begin{array}{@{}r@{\,}c@{\,}ll@{}r@{}}
\multicolumn{4}{@{}l}{\textbf{Set of traced event-handlers}} \\
T & ::= & n \rightharpoonup h \\
\multicolumn{4}{@{}l}{\textbf{Events}} \\
\mathit{ev}   & ::=  & \multicolumn{2}{@{\,}l}{\texttt{'listen'} \mid \texttt{'get'} \mid \texttt{'post'} \mid \cdots} \\
\multicolumn{4}{@{}l}{\textbf{Event Handler}} \\
h & ::= & \multicolumn{2}{@{\,}l}{\tHandler{x}{x}{t}} \\
\multicolumn{4}{@{}l}{\textbf{l-values}} \\
\tLval & ::=  & x              & \textrm{Variable} \\
\cline{1-4} \multicolumn{1}{@{}l}{\vline}
       & \mid & \tOp{*}t\tOp{.}x & \textrm{Variable in environment} & \vline \\
\multicolumn{3}{@{}l}{\vline \textbf{Addresses}} & & \vline \\
\multicolumn{1}{@{}l}{\vline \tAddr} 
      & ::=  & t\tOp{.}x & \textrm{Address in environment} & \vline \\
\multicolumn{1}{@{}l}{\vline}
      & \mid & \tOp{\&}x & \textrm{Address of variable} & \vline \\
\cline{1-4}
\multicolumn{4}{@{}l}{\textbf{Blocks}} \\
\tBlk & ::= & \tOp{\{}~t_1 \cdots t_n\tOp{\}}
\end{array}
\)
\quad\vrule\quad
\(
\begin{array}{@{}r@{\,}c@{\,}ll@{}r@{}}
\multicolumn{4}{@{}l}{\textbf{Trace trees}} \\
t & ::=  & c & \textrm{Constant} \\
  & \mid & x                & \textrm{Variable} \\
  & \mid & t_1~\mathit{op}~t_2 & \textrm{Binary operation}  \\
  & \mid & \tBlk
         & \textrm{Block} \\
  & \mid & \tKw{if}~\tOp{(}t_1\tOp{)}~t_2~\tKw{else}~t_3
         & \textrm{Conditionals} \\
  & \mid & \tKw{while}~\tOp{(}t_1\tOp{)}~\tBlk
          & \textrm{Loops} \\
  & \mid & \tKw{let}~x~\tOp{=}~t\tOp{;} & \textrm{Variable declaration} \\
  & \mid & \tLval~\tOp{=}~t\tOp{;} & \textrm{Assignment} \\
  & \mid & \ell\tOp{:} t
         & \textrm{Labelled trace} \\
  \cline{1-4} \multicolumn{1}{@{}l}{\vline}
  & \mid & \tUnk{}
  & \textrm{Unknown behavior} & \vline \\
  \multicolumn{1}{@{}l}{\vline}
  & \mid & \tKw{break}~\ell~t\tOp{;}
  & \textrm{Break with value} & \vline \\
  \multicolumn{1}{@{}l}{\vline}
  & \mid & \tEvent{\mathit{ev}}{t_{\mathit{arg}}}{t_{\mathit{env}}}{n}
  & \textrm{Event handler} & \vline \\
  \multicolumn{1}{@{}l}{\vline}
  & \mid & \tKw{respond}\tOp{(}t\tOp{)}
  & \textrm{Response} & \vline \\
  \multicolumn{1}{@{}l}{\vline}
  & \mid & \tKw{env}\tOp{(}x_1\tOp{:}\tAddr_1\tOp{,}\cdots\tOp{,}x_n\tOp{:}\tAddr_n\tOp{)}
  & \textrm{Environment object} & \vline \\
  \multicolumn{1}{@{}l}{\vline}
  & \mid & \tOp{*}t\tOp{.}x
  & \textrm{Value in environment} & \vline \\
  \cline{1-4}
\end{array}
\)

\caption{The language of traces, most of which corresponds to JavaScript
without functions. The boxed portions do not have JavaScript counterparts.}

\label{trace-ir-figure}

\end{figure}

\subsection{The Language of Traces}

\sysname{} instruments a serverless function written in JavaScript to
dynamically generate a program in a \emph{trace language}. On any
input, the trace either 1)~exhibits the same behavior as the original
JavaScript program, or 2)~halts with a fatal error that indicates
unknown behavior (\tUnk). \Cref{trace-ir-figure} shows the trace language using
syntax that resembles JavaScript. In practice, since we do not write trace
programs by hand, they do not need a human-readable syntax.\footnote{The
implementation of \sysname{} represents traces using JSON.} Many features of
the trace language correspond directly to JavaScript, which is to be expected,
since it represents a JavaScript program. However, the trace language lacks
user-defined functions, as they get eliminated during tracing
(\cref{instrumentation}). The trace language also includes several kinds of
expressions that do not correspond to JavaScript---the boxed
expressions in \cref{trace-ir-figure}--which we describe below. This paper
denotes JavaScript syntax in $\jsOp{blue}$ and the trace language syntax in
$\tOp{red}$.

\paragraph{Unknown behavior}

Since the generated trace may not cover all possible code-paths in the
serverless function, the language includes an expression that indicates unknown
behavior (\tUnk). Evaluating this expression
aborts the language-based sandbox and restarts execution in a container.

\paragraph{Unified statements and expressions}

The trace language unifies expressions and statements. For example, the
following trace, uses a loop, block, and a variable declaration in
expression position:
\[
\tKw{let}~x~\tOp{=}~\tOp{\{}~\tKw{let}~y~\tOp{=}~\tOp{\num{5}}\tOp{;}~\tKw{while}~\tOp{(}y\tOp{>}\tOp{\num{0}}\tOp{)}~\tOp{\{}~y~\tOp{=}~y~\tOp{-}\tOp{\num{1}}\tOp{;}~\tOp{\}}~y~\tOp{\}}\tOp{;}
\]
In addition, the trace language unifies JavaScript's $\jsKw{break}$ and
$\jsKw{return}$ statements into a single expression that breaks to a label and
returns a value ($\tKw{break}~\ell~t$). These choices make interprocedural
tracing significantly easier, and since Rust has a similar design, they do not
make generating Rust code any harder.

\paragraph{Explicit environment representation}

When several JavaScript functions close over a shared variable, their closure
objects contain aliases to the same memory location. Although the trace
language does not have first-class functions, it must correctly preserve this
form of aliasing. Therefore, the language includes explicit environment objects
($\tKw{env}$), which are a record of variable names and their addresses. The
trace language also has expressions to read a value from an environment
($\tOp{*}t\tOp{.}x$), read an address from an environment ($t\tOp{.}x$), and
get the address of a variable ($\tOp{\&}x$).

\paragraph{Events handlers}

To successfully trace serverless function, traces must be able to represent
asynchronous code paths, and not just sequential control. Therefore, the result
of tracing is a set of numbered event handlers ($\tKw{handler}$).
In a trace program each handler contains 1)~the body of the event
handler, which is a trace tree that runs in response to the event
($\tKw{body}$), 2)~the name of a variable that refers to the event itself
($\tKw{argId}$), 3)~the name of of a variable that refers to an environment object
($\tKw{envId}$). Thus the two aforementioned variable names may occur free in the
handler's body.

In addition, handlers have a fourth field, which is the \emph{value} of the
environment ($\tKw{env}$). This value is only available at runtime, and thus
does not appear in the syntax of a handler. The environment allows us to
support callbacks that close over variables in their environment, which
are common in JavaScript.

We assume that there is always a handler numbered $0$ that contains the trace
for the main body of the program. Therefore, to execute a program, we run the
trace tree in handler zero with a dummy argument and an empty environment.
The other event handlers do not run
until the program issues an event using the $\tKw{event}$ expression, which
requires several arguments:
\begin{enumerate}

  \item An event type ($\mathit{ev}$), which determines the kind of operation to
  perform, e.g., to send a web request or start a timer;

  \item An event argument ($t_\mathit{arg}$), which is a trace that determines, for example, the 
  URL to request or the duration of the timer;

  \item The number of an event handler ($n$) that will be called when the event completes; and

  \item The environment ($t_\mathit{env}$), which is a trace that refers to the
  environment object of the event handler. At runtime, when \sysname{}
  evaluates an $\tKw{event}$ expression, it 1)~stores the value of
  $t_\mathit{env}$ in the handler $n$, 2)~fires the event $\mathit{ev}$
  (implemented in Rust), and 3)~when the event completes, it invokes the body
  of the handler $n$.

\end{enumerate}

Tracing event handlers are a unique feature of \sysname{}, which is
driven by the fact that in typical serverless functions, all ``hot paths''
include callbacks. Without this feature, our language-based sandbox
would only support trivial serverless functions that do not interact with
external services.

\begin{figure}
\footnotesize
\(
\begin{array}{@{}r@{\,}c@{\,}ll}
\cont
& ::= & \cdot 
& \textrm{Empty context} \\
& \mid & \contcase{seq}([t_1\cdots t_{i-1}],[t_{i+1}\cdots t_n],\cont)
& \textrm{In a block, with $[t_1\cdots t_{i-1}]$ already executed.} \\
& \mid & \contcase{ifTrue}(t_1,t_2,\cont)
& \textrm{In the true branch of an \textit{if}, with condition $t_1$ and false branch $t_2$.} \\
& \mid & \contcase{ifFalse}(t_1,t_2,\cont)
& \textrm{In the false branch of an \textit{if}, with condition $t_1$ and true branch $t_2$.} \\
& \mid & \contcase{while}(t,\cont)
& \textrm{In the body of a loop, with condition $t$.} \\
& \mid & \contcase{Label}(\ell,\cont)
& \textrm{In the body of a labeled trace, with label $\ell$.} \\
& \mid & \contcase{Named}(x,\cont)
& \textrm{In the body of a named variable $x$.} \\
\end{array}
\)

\caption{A trace context identifies a position within a trace in which the
current statement is executing.}

\label{trace-contexts}

\end{figure}
  
\subsection{Trace Contexts}

The \sysname{} runtime must be able to incrementally build a trace tree, and
efficiently merge the trace of the current execution into the existing trace
tree. To make this possible, \sysname{} uses an explicit representation of
\emph{trace contexts} (\cref{trace-contexts}).

Similar to a continuation, a trace context ($\cont$) is a representation of a
trace with a ``hole''. For example, consider the following trace-with-a-hole (which
we indicate with $\Box$):
\[
\tKw{while}~\tOp{(}y\tOp{<}\tOp{\num{0}}\tOp{)}~\tOp{\{}\tKw{if}~\tOp{(}x\tOp{>}\tOp{\num{0}}\tOp{)}~\Box~\tKw{else}~\tUnk\tOp{\}}
\]
We can represent this trace-with-a-hole as the following trace context:
\[ \contcase{ifTrue}(x\tOp{>}\tOp{\num{0}}, \tUnk , \contcase{while}(y\tOp{<}\tOp{\num{0}}, \cdot)) \]
In this example, the $\contcase{ifTrue}$ indicates that the $\Box$ is
immediately inside the true-branch of the inner conditional, which is
immediately inside the loop ($\contcase{while}$), which is at the top-level
($\cdot$).

Each layer of the trace context carries enough information to completely
reconstruct the trace. Thus $\contcase{ifTrue}$ carries the trace of
the condition ($x\tOp{>}\tOp{\num{0}}$) and the false branch ($\tUnk$),
and $\contcase{while}$ carries the trace of the loop guard ($y\tOp{<}\tOp{\num{0}}$).
Notice that the trace context represents the expressions around the hole
``inside out''. This representation makes trace context manipulation simpler
and more efficient for the tracing runtime system.

Finally, we note that a trace context is \emph{not} a continuation. For
example, the continuation frame of an \textit{if} expression
($\tKw{if}~\tOp{(}t_1\tOp{)}~t_2~\tKw{else}~t_3$) is the following context:
\[\tKw{if}~\tOp{(}\Box\tOp{)}~t_2~\tKw{else}~t_3\]
This indicates that the
current expression is within the conditional. In contrast, $\contcase{ifTrue}$
is analogous to the following context: \[\tKw{if}~\tOp{(}t_1\tOp{)}~\Box~\tKw{else}~t_3\] This 
indicates that the current expression is within the true-branch.
In fact, the runtime system, which we present in the next section, uses trace
contexts to build a ``zipper'' for the program's trace.

\begin{figure}
\footnotesize
\(
\begin{array}{@{}r@{\,}c@{\,}l}
\multicolumn{3}{@{}c}{\env : x \rightarrow \astRep{t} \quad\quad \compileLVal\astRep{x}\env \triangleq \env(x) \quad\quad \compileExpr\astRep{c}\env \triangleq \astRep{c} \quad\quad \compileExpr\astRep{x}\env \triangleq \env(x)} \\
\compileExpr\astRep{\jsExpr_1~\mathit{op}~\jsExpr_2}\env & \triangleq
  & \jsExpr'_1~\astRep{\mathit{op}}~\jsExpr'_2 \quad \textrm{where}~\jsExpr'_1 \triangleq \compileExpr\astRep{\jsExpr_1}\env \quad \jsExpr'_2 \triangleq \compileExpr\astRep{\jsExpr_2}\env \\
\compileStmt\astRep{\jsKw{let}~x~\jsOp{=}~\jsExpr\jsOp{;}}\env & \triangleq 
  & (\rtsFun{let}{\astRep{x},\compileExpr\astRep{\jsExpr}\rho}\jsOp{;}{\jsKw{let}~x~\jsOp{=}~\jsExpr\jsOp{;}}, \env[x \mapsto \astRep{x}]) \\
\compileStmt\astRep{\jsKw{let}~f~\jsOp{=}~\jsKw{function}\jsOp{(}\jsParam_1\cdots\jsParam_n\jsOp{)}\jsBlk\jsOp{;}}\env & \triangleq 
  & (\rtsFun{let}{\astRep{f},\astRep{\rho}}\jsOp{;}{\jsKw{let}~f~\jsOp{=}~\jsKw{function}\jsOp{(}\jsParam_1\cdots\jsParam_n\jsOp{)}\jsBlk''\jsOp{;}}, \env[f \mapsto \astRep{f}]) \\
  & & ~\textrm{where}~\jsStmt_1 \triangleq \rtsFun{let}{\astRep{x_1}, \rtsFun{popArg}{}} ~ \cdots ~ \jsStmt_n \triangleq \rtsFun{let}{\astRep{x_n}, \rtsFun{popArg}{}} \\
  & & ~\phantom{\textrm{where}}~\jsOp{\{}\jsStmt'_1\cdots\jsStmt'_m\jsOp{\}} \triangleq \jsBlk \\
  & & ~\phantom{\textrm{where}}~\jsOp{(}y_1 \cdots y_q\jsOp{)} \triangleq \rtsFun{domain}{\env} \\
& & ~\phantom{\textrm{where}}~\env' \triangleq \env\left[\begin{array}{l}\jsParam_1 \mapsto \astRep{\jsParam_1}\cdots\jsParam_n\mapsto\astRep{\jsParam_n},\\\ensuremath{y}_1\mapsto\astRep{\tEnvid\tOp{.}\ensuremath{y}_1}\cdots\ensuremath{y}_q\mapsto\astRep{\tEnvid\tOp{.}\ensuremath{y}_q}\end{array}\right] \\
  & & ~\phantom{\textrm{where}}~(\jsBlk', \env'') \triangleq \compileStmt\astRep{\jsOp{\{}\rtsFun{let}{\astRep{\tEnvid},\rtsFun{popArg}{}}\jsOp{;}s_1 \cdots s_n\jsOp{;}s'_1 \cdots s'_m\jsOp{\}}}\env' \\
  & & ~\phantom{\textrm{where}}~\jsBlk'' \triangleq \jsOp{\{}\rtsFun{label}{\tRetlabel}\jsOp{;}\jsBlk'\jsOp{;}\rtsFun{pop}{}\jsOp{\}}\\
\compileStmt\astRep{\jsKw{let}~r~\jsOp{=}~f\jsOp{(}\jsExpr_1\cdots\jsExpr_n\jsOp{)}\jsOp{;}}\env & \triangleq 
  & (\jsStmt_n\cdots\jsStmt_1\jsOp{;}\rtsFun{pushArg}{\astRep{f}}\jsOp{;}\rtsFun{named}{\astRep{r}}\jsOp{;}\jsKw{let}~r~\jsOp{=}~f\jsOp{(}\jsExpr_1\cdots\jsExpr_n\jsOp{)}\jsOp{;}\rtsFun{pop}{}\jsOp{;}, \env') \\
  & & ~\textrm{where}~\jsStmt_1 \triangleq \rtsFun{pushArg}{\compileExpr\astRep{\jsExpr_1}\env} ~ \cdots ~ \jsStmt_n \triangleq \rtsFun{pushArg}{\compileExpr\astRep{\jsExpr_n}\env} \\  
  & & ~\phantom{\textrm{where}}~\env' \triangleq \env[r \mapsto \astRep{r}] \\
\compileStmt\astRep{\mathit{lval}~\jsOp{=}~\jsExpr\jsOp{;}}\env & \triangleq 
  & (\rtsFun{set}{\compileLVal\astRep{\mathit{lval}}\rho,\compileExpr\astRep{\jsExpr}\rho}\jsOp{;}{\mathit{lval}~\jsOp{=}~\jsExpr\jsOp{;}}, \env]) \\
\compileStmt\astRep{\jsOp{\{}\jsStmt_1\cdots\jsStmt_n\jsOp{\}}}\env & \triangleq
  & (\jsOp{\{}\rtsFun{enterSeq}{n}\jsOp{;}\jsStmt'_1\jsOp{;}\rtsFun{seqNext}{}\jsOp{;}\jsStmt'_2\jsOp{;}\cdots\jsOp{;}\jsStmt'_n\jsOp{;}\rtsFun{pop}{}\jsOp{\}},\env) \\
  & & ~\textrm{where}~(\jsStmt'_1, \env_1) \triangleq \compileStmt\astRep{\jsStmt_1}\env ~ \cdots ~ (\jsStmt'_n, \env_n) \triangleq \compileStmt\astRep{\jsStmt_n}\env_{n-1} \\
\compileStmt\astRep{\jsKw{if}~(\jsExpr)~\jsStmt_1~\jsKw{else}~\jsStmt_2}\env & \triangleq
  & (\jsKw{if}~(\jsExpr)~\jsOp{\{}\rtsFun{ifTrue}{\compileExpr\astRep{\jsExpr}\rho}\jsOp{;}\jsStmt'_1\jsOp{\}}~\jsKw{else}~\jsOp{\{}\rtsFun{ifFalse}{\compileExpr\astRep{\jsExpr}\rho}\jsOp{;}\jsStmt'_2\jsOp{\}}\jsOp{;}\rtsFun{pop}{},\env) \\
  & & ~\textrm{where}~(\jsStmt'_1,\env_1) \triangleq \compileStmt\astRep{\jsStmt_1}\env \quad (\jsStmt'_2,\env_2) \triangleq \compileStmt\astRep{\jsStmt_2}\env \\
\compileStmt\astRep{\jsKw{while}~(\jsExpr)~\jsStmt}\env & \triangleq
  & (\rtsFun{while}{\compileExpr\astRep{\jsExpr}\rho}\jsOp{;}\jsKw{while}~(\jsExpr)~\jsStmt'\jsOp{;}\rtsFun{pop}{},\env) \\
  & & ~\textrm{where}~(\jsStmt',\env') \triangleq \compileStmt\astRep{\jsStmt}\env \\
\compileStmt\astRep{\ell\jsOp{:}\jsStmt}\env & \triangleq
  & (\rtsFun{label}{\ell}\jsOp{;}\ell\jsOp{:}\jsStmt',\env) \quad \textrm{where}~(\jsStmt', \env')=\compileStmt\astRep{\jsStmt}\env \\
\compileStmt\astRep{\jsKw{break}~\ell\jsOp{;}}\env & \triangleq
  & (\rtsFun{break}{\ell,\tKw{undefined}}\jsOp{;}\rtsFun{popTo}{\ell}\jsOp{;}\jsKw{break}~\ell\jsOp{;},\env) \\
\compileStmt\astRep{\jsKw{return}~\jsExpr\jsOp{;}}\env & \triangleq
  & (\rtsFun{break}{\tRetlabel,\compileExpr\astRep{\jsExpr}\env}\jsOp{;}\rtsFun{popTo}{\tRetlabel}\jsOp{;}\jsKw{return}~\jsExpr\jsOp{;},\env) \\
\compileStmt\astRep{\astRep{t}}\env & \triangleq
  & (\astRep{t},\env) \\
\end{array}
\)

\caption{The trace compiler.}
\label{trace-compiler-figure}

\end{figure}

\begin{figure}
\footnotesize
\(
\begin{array}{@{}r@{\,}c@{\,}ll}
\multicolumn{4}{@{}l}{\textrm{\textbf{Operations that create leaves in the trace tree}}} \\ \hline
\rtsFun{let}{x, t} & \triangleq
  & c\texttt{=}\tKw{let}~x~\tOp{=}~t\tOp{;} \\
\rtsFun{set}{t_1, t_2} & \triangleq
  & c\texttt{=}t_1~\tOp{=}~t_2\tOp{;} \\
\rtsFun{break}{\ell,t} & \triangleq
  & c\texttt{=}\tKw{break}~\ell~t\tOp{;} \\
\multicolumn{4}{@{}l}{\textrm{\textbf{Operations that may create interior nodes in the trace tree}}} \\ \hline
\rtsFun{enterSeq}{n} & \triangleq 
  & c\texttt{=}\tUnk{}_1;~
    \cont\texttt{=}\contcase{seq}([],[\tUnk{}_2\cdots\tUnk{}_n],\cont)
  & \textrm{if}~c = \tUnk \\
\rtsFun{enterSeq}{n} & \triangleq 
  & c\texttt{=}t_1;~
    \cont\texttt{=}\contcase{seq}([],[t_2\cdots t_n],\cont)
  & \textrm{if}~c = \tOp{\{}t_1\cdots t_n\tOp{\}} \\
\rtsFun{seqNext}{} & \triangleq
  & c\texttt{=}t_{i+1};~
    \cont\texttt{=}\contcase{seq}([t_1\cdots t_{i-1},c],[t_{i+2}\cdots t_n],\cont)
  & \textrm{if}~\cont = \contcase{seq}([t_1\cdots t_{i-1}],[t_{i+1}\cdots t_n],\cont) \\
\rtsFun{ifTrue}{t} & \triangleq
  & c\texttt{=}\tUnk{};~
    \cont\texttt{=}\contcase{ifTrue}(t,\tUnk{},\cont)
  & \textrm{if}~c = \tUnk{} \\
\rtsFun{ifTrue}{t_1} & \triangleq
  & c\texttt{=}t_2;~
    \cont\texttt{=}\contcase{ifTrue}(t_1,t_3,\cont)
  & \textrm{if}~c =\tKw{if}~\tOp{(}t_1\tOp{)}~t_2~\tKw{else}~t_3 \\
\rtsFun{ifFalse}{t} & \triangleq
  & c\texttt{=}\tUnk{};~
    \cont\texttt{=}\contcase{ifFalse}(t,\tUnk{},\cont)
  & \textrm{if}~c = \tUnk{} \\
\rtsFun{ifFalse}{t_1} & \triangleq
  & c\texttt{=}t_3;~
    \cont\texttt{=}\contcase{ifFalse}(t_1,t_2,\cont)
  & \textrm{if}~c =\tKw{if}~\tOp{(}t_1\tOp{)}~t_2~\tKw{else}~t_3 \\
\rtsFun{while}{t} & \triangleq
  & c\texttt{=}\tUnk{};~
    \cont\texttt{=}\contcase{while}(t,\cont)
  & \textrm{if}~c = \tUnk{} \\
\rtsFun{while}{t_1} & \triangleq
  & c\texttt{=}t_2;~
    \cont\texttt{=}\contcase{while}(t_1,\cont)
  & \textrm{if}~c =\tKw{while}~\tOp{(}t_1\tOp{)}~t_2 \\
\rtsFun{label}{\ell} & \triangleq
  & c\texttt{=}\tUnk;~
    \cont\texttt{=}\contcase{Label}(\ell,\cont)
  & \textrm{if}~c = \tUnk{} \\
\rtsFun{label}{\ell} & \triangleq
  & c\texttt{=}t;~
    \cont\texttt{=}\contcase{Label}(\ell,\cont)
  & \textrm{if}~c = \ell\tKw{:} t \\
\rtsFun{named}{x} & \triangleq
  & c\texttt{=}\tUnk\tOp{;}
    \cont\texttt{=}\contcase{Named}(x,\cont)
  & \textrm{if}~c = \tUnk{} \\
\rtsFun{named}{x} & \triangleq
  & c\texttt{=}t;~
    \cont\texttt{=}\contcase{Named}(x,\cont)
  & \textrm{if}~c = \tKw{let}~x~\tOp{=}~t \\
\multicolumn{4}{@{}l}{\textrm{\textbf{Operations that move from a node to its parent in the trace tree}}} \\ \hline
\rtsFun{pop}{} & \triangleq
  & c\texttt{=}\tKw{if}~\tOp{(}t_1\tOp{)}~c~\tKw{else}~t_2;~
    \cont\texttt{=}\cont'
  & \textrm{if}~\cont = \contcase{ifTrue}(t_1,t_2,\cont') \\
\rtsFun{pop}{} & \triangleq
  & c\texttt{=}\tKw{if}~\tOp{(}t_1\tOp{)}~t_2~\tKw{else}~c;~
    \cont\cont'
  & \textrm{if}~\cont = \contcase{ifFalse}(t_1,t_2,\cont') \\
\rtsFun{pop}{} & \triangleq
  & c\texttt{=}\tKw{while}~\tOp{(}t\tOp{)}~c;~
    \cont\texttt{=}\cont'
  & \textrm{if}~\cont = \contcase{while}(t,\cont') \\
\rtsFun{pop}{} & \triangleq
  & c\texttt{=}\tOp{\{}t_1\cdots t_{i-1}\tOp{;}c\tOp{;}t_{i+1}\cdots t_n\tOp{\}};~
    \cont\texttt{=}\cont'
  & \textrm{if}~\cont = \contcase{seq}([t_1\cdots t_{i-1}],[t_{i+1}\cdots t_n],\cont') \\
\rtsFun{pop}{} & \triangleq
  & c\texttt{=}\ell\tOp{:}~c;~
    \cont\texttt{=}\cont'
  & \textrm{if}~\cont = \contcase{Label}(\ell,\cont') \\
\rtsFun{pop}{} & \triangleq
  & c\texttt{=}\tKw{let}~x~\tOp{=}~c;~
    \cont\texttt{=}\cont'
  & \textrm{if}~\cont = \contcase{Named}(x,\cont') \\
\rtsFun{popTo}{\ell} & \triangleq
  & c\texttt{=}\ell\tOp{:}~c;~
    \cont\texttt{=}\cont'
  & \textrm{if}~\cont = \contcase{Label}{(\ell,\cont')} \\
\rtsFun{popTo}{\ell} & \triangleq
  & \rtsFun{pop}{}\jsOp{;}~\rtsFun{popTo}{\ell}\jsOp{;}
  & \textrm{if}~\cont \ne \contcase{Label}{(\ell,\cont')} \\
\multicolumn{4}{@{}l}{\textrm{\textbf{Operations that manipulate the stack of argument traces}}} \\ \hline
\rtsFun{pushArg}{t} & \triangleq
  & \alpha\texttt{=}(t::\alpha) \\
\rtsFun{popArg}{} & \triangleq
  & \alpha\texttt{=}\alpha';~\jsKw{return}~t\jsOp{;}
  & \textrm{if}~\alpha~=~(t::\alpha) \\
\end{array}
\)

\caption{The functions provided by the tracing runtime system. We initialize $c
= \tUnk$, $\cont = \cdot$, and $\alpha = []$.}

\label{tracing-rts}

\end{figure}

\subsection{Instrumenting JavaScript to Generate Traces}
\label{instrumentation}

\sysname{} uses a source-to-source compiler (\cref{trace-compiler-figure}) to
instrument a serverless function to build its own trace. The compiler is
syntax-directed and relies on a small runtime system (\cref{tracing-rts}) to
incrementally merge the current execution trace with an existing trace tree.
The runtime system also includes functions to register event handlers that
support trace generation, which we present in \cref{event-handlers}. This
section focuses on tracing JavaScript programs that do not use callbacks.

Although the tracing runtime is a JavaScript library, this is an inessential
detail, thus we present it more abstractly. The internal state of the runtime
system consists of three variables: 1)~the trace of the currently executing
statement ($c$), 2)~its trace context ($\cont$), and 3)~a stack of traces
that represent function arguments ($\alpha$). A key invariant during tracing is
that plugging $c$ into $\cont$ produces a trace for the entire program.
Therefore, when tracing begins, we initialize $c$ to the unknown statement
($\tUnk$), $\cont$ to the empty trace context ($\cdot$), and $\alpha$ to an
empty stack $[]$.

The runtime system provides several functions that manipulate $c$, $\cont$, and
$\alpha$. The compiler produces a JavaScript program that calls the
aforementioned functions. We write $\astRep{t}$ to denote the runtime
representation of the expression $t$. For example, $\astRep{x}$ evaluates to a
representation of the identifier $x$, whereas $x$ evaluates to its value. Most
functions in the runtime system receive runtime representations of expressions
($\astRep{t}$-arguments).\footnote{In our implementation, $\astRep{t}$ is a JSON
data structure.}

\paragraph{The tracing runtime}

The runtime system has four classes of functions:

\begin{enumerate}

  \item Several functions record an operation in the current trace, but
  leave the trace context unchanged. These functions correspond to JavaScript
  statements that do not affect the control-flow of the program, such as
  declaring a variable (\rtsFunName{let}) or assigning to a variable
  (\rtsFunName{set}). If we think of the trace expression as a tree, these
  functions create leaf nodes in the expression tree.

  \item Several functions push a new frame onto the trace context. The compiler
  inserts calls to these functions to record the control-flow of the program.
  Each function in this category has two cases. 1)~If $c$ is $\tUnk$, it
  creates a new context frame and leave the current expression as $\tUnk$. If
  we think of the trace expression as a tree, this case occurs when we enter a
  node in the trace tree for the first time.
  2)~If $c$ is not $\tUnk$, it uses the
  sub-expressions of $c$ to create the context frame and update $c$ itself.
  For example, if $c$ is a \textit{if} expression, \rtsFunName{ifTrue} 
  stores the condition and false-part in the trace context, and
  sets $c$ to the true-part. Conversely, \rtsFunName{ifFalse} sets $c$ to the
  false-part. Thinking of the trace expression as a tree, this case occurs when we
  descend into a branch of a node that we have visited before, while preserving
  other branches in the trace context.

  \item The function $\rtsFunName{pop}$ pops the top of the trace context, and
  uses it to update $c$ to a new expression, which uses the previous value of
  $c$ as a sub-expression. Thinking of the trace expression as a tree, we call
  pop to ascend from a node to its parent. We use $\rtsFunName{popTo}$ function
  to trace \tKw{break} expressions, which transfer control out of a labeled
  block. This function calls $\rtsFunName{pop}$ repeatedly until it reaches a
  block with the desired label.

  \item The functions $\rtsFunName{pushArg}$ and $\rtsFunName{popArg}$ push and
  pop traced expressions onto the stack of arguments ($\alpha$).
\end{enumerate}

Note that the current trace and its context effectively form a
``zipper''~\cite{huet:zipper} for a trace of the entire program, and the
functions defined above are closely related to canonical zipper operation. The
operations that create trace context frames are unconventional because they
either move the focus of the zipper into an existing child node, or create a
new child and then focus on it. Although we are using a zipper-like data
structure, note that the runtime system is stateful: the functions update $c$,
$\cont$, and $\alpha$. Instead, the zipper-based approach is a clean
abstraction for building the trace tree incrementally.

\paragraph{The tracing compiler}

The compiler (\cref{trace-compiler-figure}) is syntax-directed compiler and
three functions to compile statements (\compileStmt), expressions
(\compileExpr), and l-values (\compileLVal). The compiler leaves the original
program unchanged, and only inserts calls to the runtime system so that
program execution builds a trace as a side-effect.

Compiling function declarations and applications requires the most work, since
traces do not have functions. Therefore, the trace of a function application
effectively inlines the trace of the function body. The compiler takes care to
ensure that the traces correctly captures the semantics of JavaScript.

\begin{enumerate}
  
\item The compiler ensures that the trace of the function body can refer to its
actual arguments, even if function application refers to a variables that is
not in scope in the function body.

\item The compiler ensures that the trace of the function body can refer to
variables that were in scope in the original JavaScript program, but are not in
scope at the application site. For this to work, the compiler represents the
trace of a function $f$ as its environment ($\rho$), function applications
pass the environment on the trace argument stack, and we bind free variables
in the function body to expressions that access fields of the environment.

\end{enumerate}

Finally, we note that function applications rely on \textit{let} expressions
(not statements) in the trace language. Before entering the body of a function,
the application uses the runtime function $\rtsFunName{named}$ to create a
trace context that encloses the named expression of a \textit{let}. We
evaluate the body of the function within this context, thus the named
expression may contain arbitrary nested expressions.

\begin{figure}

\begin{subfigure}{0.325\columnwidth}
\footnotesize
\(
\def\arraystretch{1.858}
\begin{array}{@{}l}
\textrm{Tracing state} \\
\hline
c = \tUnk\tikzmark{if1-mark1}, \cont = \cdot \\
\hline
c = \tUnk, \cont = \contcase{ifTrue}(x\tOp{<}\tOp{\num{0}},\tikzmark{if1-mark2}\tUnk,\cdot) \tikzmark{ex-if-ifTrue-cfg}  \\
\hline
c = y\tOp{=}x\tOp{*}\tOp{\num{-1}}\tOp{;}, \cont = \contcase{ifTrue}(x\tOp{<}\tOp{\num{0}},\tikzmark{if1-mark3}\tUnk,\cdot) \tikzmark{ex-if-ifTrue-set-cfg}  \\
\hline
c = \tKw{if}\tOp{(}x\tOp{<}\tOp{\num{0}}\tOp{)}y\tOp{=}x\tOp{*}\tOp{\num{-1}}\tOp{;}\tKw{else}\tUnk\tikzmark{if1-mark4}, \cont = \cdot \tikzmark{ex-if-pop-cfg-1} \\
\end{array}
\)
  
\caption{First execution.}
\label{if-example-output-1}

\end{subfigure}
\vrule
\begin{subfigure}{0.24\columnwidth}
\lstset{language=JavaScript}
\begin{lstlisting}[
  numbers=none,
  numbersep=0pt,
  resetmargins=true,
  linebackgroundcolor={%
  \ifnum\value{lstnumber}=2\color{gray!30}\fi
  \ifnum\value{lstnumber}=3\color{gray!30}\fi
  \ifnum\value{lstnumber}=6\color{gray!30}\fi
  \ifnum\value{lstnumber}=7\color{gray!30}\fi
  \ifnum\value{lstnumber}=10\color{gray!30}\fi
  }]
#\jsKw{if}~\jsOp{(}x\jsOp{<}\jsOp{0}\jsOp{)}~\jsOp{\{}#
  #\tikzmark{ex-if-ifTrue-lhs}\rtsFun{ifTrue}{\llbracket x\tOp{<}\tOp{\num{0}}\rrbracket}\jsOp{;}#
  #\tikzmark{ex-if-ifTrue-set-lhs}\rtsFun{set}{\llbracket y \rrbracket, \llbracket x\tOp{*}\tOp{\num{-1}}\rrbracket}\jsOp{;}#
  #y~\jsOp{=}~x~\jsOp{*}~\jsOp{-1}\jsOp{;}#
#\jsOp{\}}~\jsKw{else}~\jsOp{\{}#
  #\rtsFun{ifFalse}{\llbracket x\tOp{<}\tOp{\num{0}}\rrbracket}\jsOp{;}##\tikzmark{ex-if-ifFalse-rhs}#
  #\rtsFun{set}{\llbracket y \rrbracket, \llbracket x\rrbracket}\jsOp{;}##\tikzmark{ex-if-ifFalse-set-rhs}#
  #y~\jsOp{=}~x\jsOp{;}#
#\jsOp{\}}#
#\tikzmark{ex-if-pop-lhs}\rtsFun{pop}{}\jsOp{;}##\tikzmark{ex-if-pop-rhs}#
\end{lstlisting}

\caption{Program.}
\label{if-example-traced}


\end{subfigure}
\vrule\,\,
\begin{subfigure}{0.34\columnwidth}
\footnotesize
\(
\def\arraystretch{1.858}
\begin{array}{@{}l}
\textrm{Tracing state} \\
\hline
c = \tKw{if}\tOp{(}x\tOp{<}\tOp{\num{0}}\tOp{)}y\tOp{=}x\tOp{*}\tOp{\num{-1}}\tKw{else}\tikzmark{if2-mark1}\tUnk,
\cont = \cdot \\
\hline
\tikzmark{ex-if-ifFalse-cfg}c = \tUnk\tikzmark{if2-mark2},
\cont = \contcase{ifFalse}(x\tOp{<}\tOp{\num{0}},y\tOp{=}x\tOp{*}\tOp{\num{-1}},\cdot)  \\
\hline
\tikzmark{ex-if-ifFalse-set-cfg}c = y\tOp{=}x;,
\cont = \contcase{ifFalse}(x\tOp{<}\tOp{\num{0}},y\tOp{=}x\tOp{*}\tOp{\num{-1}},\cdot) \\
\hline
\tikzmark{ex-if-pop-cfg-2}c = \tKw{if}\tOp{(}x\tOp{<}\tOp{\num{0}}\tOp{)}y\tOp{=}x\tOp{*}\tOp{\num{-1}}\tOp{;}\tKw{else}y\tOp{=}x\tOp{;},
\cont = \cdot \\
\end{array}
\)
    
\caption{Second execution.}
\label{if-example-output-2}

\end{subfigure}

\begin{tikzpicture}[remember picture,overlay]
\draw[gray,-latex] (pic cs:if2-mark1) to [bend left=30] ([yshift=5]pic cs:if2-mark2);
\draw[gray,-latex] ([yshift=5]pic cs:if1-mark1) to [bend left=30] ([yshift=5]pic cs:if1-mark2);
\draw[gray,-latex] ([yshift=-3,xshift=4]pic cs:if1-mark3) to [bend left=30] ([yshift=5]pic cs:if1-mark4);
\draw[dashed,-latex](pic cs:ex-if-ifTrue-lhs) to (pic cs:ex-if-ifTrue-cfg);
\draw[dashed,-latex](pic cs:ex-if-ifTrue-set-lhs) to (pic cs:ex-if-ifTrue-set-cfg);
\draw[dashed,-latex](pic cs:ex-if-pop-lhs) to (pic cs:ex-if-pop-cfg-1);
\draw[dashed,-latex](pic cs:ex-if-ifFalse-rhs) to (pic cs:ex-if-ifFalse-cfg);
\draw[dashed,-latex](pic cs:ex-if-ifFalse-set-rhs) to (pic cs:ex-if-ifFalse-set-cfg);
\draw[dashed,-latex](pic cs:ex-if-pop-rhs) to (pic cs:ex-if-pop-cfg-2);
\end{tikzpicture}

\caption{An example of incremental trace tree construction.
\Cref{if-example-traced} shows the JavaScript program, where the shaded lines
are those that the compiler inserts and the unshaded lines are those that were
present in the original program. \Cref{if-example-output-1} shows the result
tracing on an initial input with $x < 0$. \Cref{if-example-output-1} shows the
result tracing on a second input with $x \ge 0$.}

\label{if-example}

\end{figure}
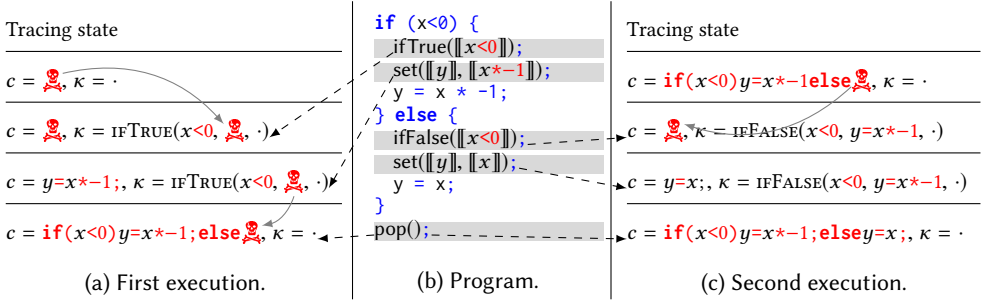
  
\paragraph{Example: tracing a conditional}

\Cref{if-example} shows an example of how the tracing compiler and runtime
system operate. The program in \cref{if-example-traced} sets $y$ to the
absolute value of $x$. To do so, it branches on $x\jsOp{<}\jsOp{\num{0}}$ and
thus has two control-flow paths. The figure shows the output of the compiler,
with the generated code shaded gray, and the original program unshaded.

\Cref{if-example-output-1} shows a first run of the program with
$x\jsOp{<}\jsOp{\num{0}}$. The initial value of the current trace ($c$) is
unknown ($\tUnk$) and the initial trace context is empty ($\cdot$). Since
$x\jsOp{<}\jsOp{\num{0}}$, the program enters the true-branch, and calls the
function \rtsFunName{ifTrue} in the runtime system. This function pushes an
\contcase{IfTrue} frame onto the trace context, that records the condition
($x\tOp{<}\tOp{\num{0}}$) and has to use $\tUnk$ as the trace of the
false-branch, since it has not been executed. After the call to
\rtsFunName{ifTrue}, the JavaScript code assigns to $y$, and the inserted call
to \rtsFunName{set} replaces the current trace (which is $\tUnk$) with a
corresponding assignment to $y$ in the trace language.
Finally, after the \textit{if} statement, the program calls \rtsFunName{pop},
which pops the \contcase{IfTrue} frame off the trace context and combined
it with the current trace---of the true-branch---to construct a trace
of the \text{if} statement. Since this is final configuration, the trace context
is empty and the current trace represents the entire known program, which
includes a $\tUnk$ in the false branch.

\Cref{if-example-output-2} shows a second run of the program with
$x\color{blue}\ge\color{black}\jsOp{\num{0}}$. This run resumes tracing where
the first run ended, thus we preserve the value of the current trace.
\footnote{The trace context is guaranteed to be empty at the start of each run.}
Since $x\color{blue}\ge\color{black}\jsOp{\num{0}}$, the program enters
the false branch and calls \rtsFunName{ifFalse}. This function is symmetric to
the \rtsFunName{ifTrue}. However, since the
current trace is already an \textit{if} expression, \rtsFunName{ifFalse} 
pushes an \contcase{IfFalse} frame onto the trace context that preserves
trace of the true branch that we calculated on the first run.
After the call to \rtsFunName{ifFalse}, the program assigns to $y$ and records
the assignment in $c$, similar to the first run.
Therefore, when the program finally calls \rtsFunName{pop}, $c$ contains
a complete trace of the false branch, and the  \contcase{IfFalse} frame
contains a complete trace of the true branch (from the prior run). Therefore,
the final value of $c$ is a complete trace without any $\tUnk$s.

\begin{figure}

\begin{subfigure}{0.32\columnwidth}
\vspace{80pt}
\lstset{language=JavaScript}
\begin{lstlisting}[
  numbers=none,
  numbersep=0pt,
  resetmargins=true,
  linebackgroundcolor={%
  \ifnum\value{lstnumber}=1\color{gray!30}\fi
  \ifnum\value{lstnumber}=2\color{gray!30}\fi
  \ifnum\value{lstnumber}=4\color{gray!30}\fi
  \ifnum\value{lstnumber}=5\color{gray!30}\fi
  \ifnum\value{lstnumber}=7\color{gray!30}\fi
  \ifnum\value{lstnumber}=8\color{gray!30}\fi
  \ifnum\value{lstnumber}=9\color{gray!30}\fi
  \ifnum\value{lstnumber}=10\color{gray!30}\fi
  \ifnum\value{lstnumber}=11\color{gray!30}\fi
  \ifnum\value{lstnumber}=12\color{gray!30}\fi
  \ifnum\value{lstnumber}=13\color{gray!30}\fi
  \ifnum\value{lstnumber}=14\color{gray!30}\fi
  \ifnum\value{lstnumber}=16\color{gray!30}\fi
  \ifnum\value{lstnumber}=17\color{gray!30}\fi
  \ifnum\value{lstnumber}=19\color{gray!30}\fi
  \ifnum\value{lstnumber}=20\color{gray!30}\fi
  \ifnum\value{lstnumber}=21\color{gray!30}\fi
  \ifnum\value{lstnumber}=22\color{gray!30}\fi
  \ifnum\value{lstnumber}=24\color{gray!30}\fi
  \ifnum\value{lstnumber}=25\color{gray!30}\fi
  }]
#\rtsFun{enterSeq}{3}\jsOp{;}\tikzmark{F-line1-js}#
#\rtsFun{let}{\llbracket x \rrbracket, \llbracket \tOp{\num{10}} \rrbracket}\jsOp{;}\tikzmark{F-line2-js}#
#\jsKw{let}~x~\jsOp{=}~\jsOp{\num{10}}\jsOp{;}#
#\rtsFun{seqNext}{}\jsOp{;}\tikzmark{F-line4-js}#
#\rtsFun{let}{\llbracket F \rrbracket, \llbracket \tKw{env}\tOp{(}x~\tOp{:}~\llbracket x \rrbracket\tOp{)}\rrbracket}\jsOp{;}\tikzmark{F-line5-js}#
#\jsKw{let}~F~\jsOp{=}~\jsKw{function}\jsOp{(}y\jsOp{)}~\jsOp{\{}#
  #\rtsFun{label}{\tRetlabel}\jsOp{;}\tikzmark{F-line7-js}#
  #\jsOp{\{}~\rtsFun{enterSeq}{3}\jsOp{;}\tikzmark{F-line8-js}#
    #\rtsFun{let}{\llbracket \tEnvid \rrbracket, \rtsFun{popArg}{}}\jsOp{;}\tikzmark{F-line9-js}#
    #\rtsFun{seqNext}{}\jsOp{;}\tikzmark{F-line10-js}#
    #\rtsFun{let}{\llbracket y \rrbracket, \rtsFun{popArg}{}}\jsOp{;}\tikzmark{F-line11-js}#
    #\rtsFun{seqNext}{}\jsOp{;}\tikzmark{F-line12-js}#
    #\rtsFun{break}{\tRetlabel, \llbracket \tEnvid\tOp{.}x\tOp{+}y \rrbracket}\jsOp{;}\tikzmark{F-line13-js}#
    #\rtsFun{popTo}{\tRetlabel}\jsOp{;}\tikzmark{F-line14-js}#
    #\jsKw{return}~x\jsOp{+}y\jsOp{;}#
    #\rtsFun{pop}{}\jsOp{;}~\jsOp{\}}\jsOp{;}\tikzmark{F-line16-js}#
  #\rtsFun{pop}{}\jsOp{;}\tikzmark{F-line17-js}#
#\jsOp{\}}\jsOp{;}#
#\rtsFun{seqNext}{}\jsOp{;}\tikzmark{F-line19-js}#
#\rtsFun{pushArg}{\llbracket \tOp{\num{3}} \rrbracket}\jsOp{;}\tikzmark{F-line20-js}#
#\rtsFun{pushArg}{\llbracket F \rrbracket}\jsOp{;}\tikzmark{F-line21-js}#
#\rtsFun{named}{\llbracket \mathit{foo} \rrbracket}\jsOp{;}\tikzmark{F-line22-js}#
#\jsKw{let}~foo~\jsOp{=}~F\jsOp{(\num{3});}#
#\rtsFun{pop}{}\jsOp{;}\tikzmark{F-line24-js}#
#\rtsFun{pop}{}\jsOp{;}\tikzmark{F-line25-js}#
\end{lstlisting}
\vspace{95pt}
\caption{Program.}
\label{fun-example-js}

\end{subfigure}
\vrule\,\,
\begin{subfigure}{0.65\columnwidth}
\footnotesize
\(
\begin{array}{@{}l}
\textrm{Tracing state}\\
\hline
\tikzmark{F-iter1-trace} c = \tUnk  \\
\alpha = [] \\
\cont = \cdot \\
\hline
\tikzmark{F-iter2-trace} c = \tKw{let}~x~\tOp{=}~\tOp{\num{10}}\tOp{;} \\
\alpha = [] \\
\cont = \contcase{seq}([], [\tUnk_2, \tUnk_3], \cdot) \\
\hline
\tikzmark{F-iter3-trace} c = \tKw{let}~F~\tOp{=}~\tKw{env}\tOp{(}x\tOp{:}x\tOp{)}\tOp{;} \\
\alpha = [] \\
\cont = \contcase{seq}([\tKw{let}~x~\tOp{=}~\tOp{\num{10}}\tOp{;}], [\tUnk_3], \cdot) \\
\hline
\tikzmark{F-iter4-trace} c = \tUnk_3 \\
\alpha = [F, \tOp{\num{3}}] \\
\cont = \contcase{seq}([\tKw{let}~x~\tOp{=}~\tOp{\num{10}}\tOp{;}\tKw{let}~F~\tOp{=}~\tKw{env}\tOp{(}x\tOp{:}x\tOp{)}\tOp{;}], [], \cdot) \\
\hline
\tikzmark{F-iter5-trace} c = \tUnk_3 \\
\alpha = [F, \tOp{\num{3}}] \\
\cont = \contcase{label}(\tRetlabel, \contcase{named}(\mathit{foo}, \contcase{seq}([\tKw{let}~x~\tOp{=}~\tOp{\num{10}}\tOp{;}\tKw{let}~F~\tOp{=}~\tKw{env}\tOp{(}x\tOp{:}x\tOp{)}\tOp{;}], [], \cdot))) \\
\hline
\tikzmark{F-iter6-trace} c = \tKw{let}~\tEnvid~\tOp{=}~F\tOp{;} \\
\alpha = [\tOp{\num{3}}] \\
\cont = \contcase{seq}([], [\tUnk'_2, \tUnk'_3], \\
\phantom{\cont = \contcase{seq}}\contcase{label}(\tRetlabel, \contcase{named}(\mathit{foo}, \contcase{seq}([\tKw{let}~x~\tOp{=}~\tOp{\num{10}}\tOp{;}\tKw{let}~F~\tOp{=}~\tKw{env}\tOp{(}x\tOp{:}x\tOp{)}\tOp{;}], [], \cdot)))) \\
\hline
\tikzmark{F-iter7-trace} c = \tKw{let}~y~\tOp{=}~\tOp{\num{3}}\tOp{;} \\
\alpha = [] \\
\cont = \contcase{seq}([\tKw{let}~\tEnvid~\tOp{=}~F\tOp{;}], [\tUnk'_3], \\
\phantom{\cont = \contcase{seq}}\contcase{label}(\tRetlabel, \contcase{named}(\mathit{foo}, \contcase{seq}([\tKw{let}~x~\tOp{=}~\tOp{\num{10}}\tOp{;}\tKw{let}~F~\tOp{=}~\tKw{env}\tOp{(}x\tOp{:}x\tOp{)}\tOp{;}], [], \cdot)))) \\
\hline
\tikzmark{F-iter8-trace} c = \tKw{break}~\tRetlabel~\tOp{(}\tEnvid\tOp{.}x\tOp{+}y\tOp{)}\tOp{;} \\
\alpha = [] \\
\cont = \contcase{seq}([\tKw{let}~\tEnvid~\tOp{=}~F\tOp{;}~\tKw{let}~y~\tOp{=}~\tOp{\num{3}}\tOp{;}], [], \\
\phantom{\cont = \contcase{seq}}\contcase{label}(\tRetlabel, \contcase{named}(\mathit{foo}, \contcase{seq}([\tKw{let}~x~\tOp{=}~\tOp{\num{10}}\tOp{;}\tKw{let}~F~\tOp{=}~\tKw{env}\tOp{(}x\tOp{:}x\tOp{)}\tOp{;}], [], \cdot)))) \\
\hline
\tikzmark{F-iter9-trace} c = \tRetlabel~\tOp{:}~\tOp{\{}\tKw{let}~\tEnvid~\tOp{=}~F\tOp{;}~\tKw{let}~y~\tOp{=}~\tOp{\num{3}}\tOp{;}\tKw{break}~\tRetlabel~\tOp{(}\tEnvid\tOp{.}x\tOp{+}y\tOp{)}\tOp{;}\tOp{\}} \\
\alpha = [] \\
\cont = \contcase{named}(\mathit{foo}, \contcase{seq}([\tKw{let}~x~\tOp{=}~\tOp{\num{10}}\tOp{;}~\tKw{let}~F~\tOp{=}~\tKw{env}\tOp{(}x\tOp{:}x\tOp{)}\tOp{;}], [], \cdot)) \\
\hline
\tikzmark{F-iter10-trace} c = \tKw{let}~\mathit{foo}~\tOp{=}~\tRetlabel~\tOp{:}~\tOp{\{}~\tKw{let}~\tEnvid~\tOp{=}~F\tOp{;}~\tKw{let}~y~\tOp{=}~\tOp{\num{3}}\tOp{;}\tKw{break}~\tRetlabel~\tOp{(}\tEnvid\tOp{.}x\tOp{+}y\tOp{)}\tOp{;}~\tOp{\}} \\
\alpha = [] \\
\cont = \contcase{seq}([\tKw{let}~x~\tOp{=}~\tOp{\num{10}}\tOp{;}\tKw{let}~F~\tOp{=}~\tKw{env}\tOp{(}x\tOp{:}x\tOp{)}\tOp{;}], [], \cdot) \\
\hline
\tikzmark{F-iter11-trace} c = \tOp{\{}\tKw{let}~x~\tOp{=}~\tOp{\num{10}}\tOp{;} \\
\phantom{c = \tOp{\{}}\tKw{let}~F~\tOp{=}~\tKw{env}\tOp{(}x\tOp{:}x\tOp{)}\tOp{;} \\
\phantom{c = \tOp{\{}}\tKw{let}~\mathit{foo}~\tOp{=}~\tRetlabel~\tOp{:}~\tOp{\{}~\tKw{let}~\tEnvid~\tOp{=}~F\tOp{;}~\tKw{let}~y~\tOp{=}~\tOp{\num{3}}\tOp{;}\tKw{break}~\tRetlabel~\tOp{(}\tEnvid\tOp{.}x\tOp{+}y\tOp{)}\tOp{;}~\tOp{\}} \\
\alpha = [] \\
\cont = \cdot \\
\end{array}
\)
\caption{Execution trace.}
\label{fun-example-trace}

\end{subfigure}

\begin{tikzpicture}[remember picture,overlay]
\draw[dashed,-latex](pic cs:F-line2-js) to (pic cs:F-iter2-trace);
\draw[dashed,-latex](pic cs:F-line5-js) to (pic cs:F-iter3-trace);
\draw[dashed,-latex](pic cs:F-line21-js) to [bend right=10] (pic cs:F-iter4-trace);
\draw[dashed,-latex](pic cs:F-line7-js) to (pic cs:F-iter5-trace);
\draw[dashed,-latex](pic cs:F-line9-js) to (pic cs:F-iter6-trace);
\draw[dashed,-latex](pic cs:F-line11-js) to [bend left=10] (pic cs:F-iter7-trace);
\draw[dashed,-latex](pic cs:F-line13-js) to (pic cs:F-iter8-trace);
\draw[dashed,-latex](pic cs:F-line14-js) to [bend left=20] (pic cs:F-iter9-trace);
\draw[dashed,-latex](pic cs:F-line24-js) to [bend right=5] (pic cs:F-iter10-trace);
\draw[dashed,-latex](pic cs:F-line25-js) to (pic cs:F-iter11-trace);
\end{tikzpicture}

\caption{An example of tracing a function application. In \cref{fun-example-js},
the unshaded lines are the original program and the shaded lines are those that
the trace compiler inserts.}

\label{fun-example}

\end{figure}
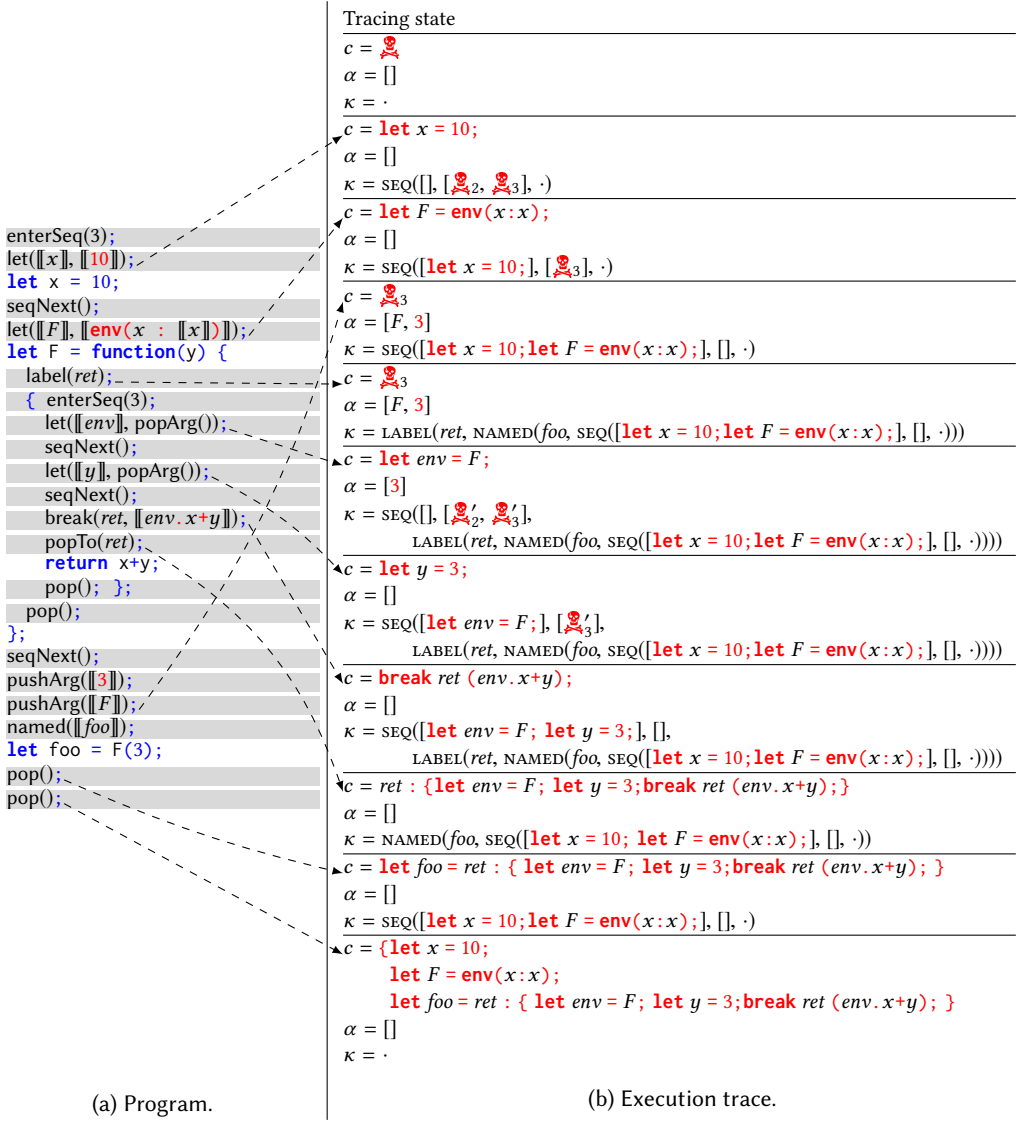
  
\paragraph{Example: tracing a function application}

\Cref{fun-example} shows an example of tracing a function application, where
the function $F(y)$ calculates $x\jsOp{+}y$ and $x$ is a free variable in the body of
$F$. \Cref{fun-example-js} shows the output of the trace compiler. As in the
previous example, the unshaded lines are the original program and the shaded
lines are those that are inserted by the compiler. \Cref{fun-example-trace} shows
the state of the tracing runtime at several points of interest. At the top
of the program, the current trace is $\tUnk$, and at the end, the trace in $c$
represents the entire program with $F$ inlined. The trace shows several
significant features of tracing:
\begin{enumerate}

  \item The trace variable \tKw{F} is bound to an trace environment that is
  equivalent to the environment of the JavaScript function named \lstinline|F|.

  \item The program pushes and pops trace expressions from the argument
  stack ($\alpha$).

  \item The runtime system uses \rtsFunName{popTo} before the \jsKw{return}
  statement, which pops frames off the trace context.

\end{enumerate}

\begin{figure}
\footnotesize
\(
\begin{array}{@{}r@{\,}c@{\,}l}
\rtsFun{newHandler}{\textit{ev}, t_\mathit{arg},t_\mathit{env}} & \triangleq
  & c\jsOp{=}\tEvent{\textit{ev}}{t_\mathit{arg}}{t_\mathit{env}}{n}\jsOp{;}
  T\jsOp{=}T[n\mapsto\tHandler{\tEnvid}{x}{\tUnk}]\jsOp{;} \\
  & & \jsKw{return}~n\jsOp{;} \quad\quad \textrm{where}~n,x~\textrm{are fresh} \quad \textrm{if}~$c$=\tUnk \\ 
\rtsFun{newHandler}{\textit{ev}, t_\mathit{arg},t_\mathit{env}} & \triangleq
  & \jsKw{return}~n\jsOp{;}  \\
  & & \textrm{if}~c\jsOp{=}\tEvent{\textit{ev}}{t_\mathit{arg}}{t_\mathit{env}}{n} \quad
  T(n) = \tHandler{\tEnvid}{x}{\tUnk}
  \\ 
\rtsFun{loadHandler}{n} & \triangleq &
  \rtsFun{pushArg}{h\tOp{.}\tKw{env}}\jsOp{;}
  \rtsFun{pushArg}{h\tOp{.}\tKw{argId}}\jsOp{;}
  c\jsOp{=}h\tOp{.}\tKw{body}\jsOp{;} 
  \quad h = T(n) \\
\rtsFun{saveHandler}{n} & \triangleq
  & T\jsOp{=}T[n\mapsto T(n)~\textrm{with}~\tKw{body} = c]\jsOp{;}
\end{array}
\)
\caption{Runtime system for event handlers.}
\label{tracing-rts-events}

\end{figure}

\begin{figure}
\begin{minipage}{0.43\columnwidth}
\begin{subfigure}{\columnwidth}
\begin{lstlisting}
#\jsKw{function}~\jsOp{get}\jsOp{(}uri\jsOp{,}~cb\jsOp{)}~\jsOp{\{}#
  #\rtsFun{popArg}{}\jsOp{;}#
  #\jsKw{let}~tUri~\jsOp{=}~\rtsFun{popArg}{}\jsOp{;}#
  #\jsKw{let}~tCb~\jsOp{=}~\rtsFun{popArg}{}\jsOp{;}#
  #\jsKw{let}~n~\jsOp{=}~\rtsFunName{newHandler}(\jsOp{'get'}, tUri, tCb)\jsOp{;}\tikzmark{ex-handler-reg}#
  #\jsKw{request}\jsOp{.get(}uri\jsOp{,}~\jsOp{(}e\jsOp{,}~resp\jsOp{)}~\jsOp{=>}~\jsOp{\{}\label{line:ex-actual-get}#
    #\rtsFunName{loadHandler}(n)\jsOp{;}\tikzmark{ex-handler-call-cb}#
    #cb\jsOp{(}resp\jsOp{.}body\jsOp{);}#
    #\rtsFunName{saveHandler}(n)\jsOp{;}\tikzmark{ex-handler-save}#
  #\jsOp{);}#
#\jsOp{\}};#
#\rtsFunName{let}($\astRep{\mathit{get}}$\jsOp{,}~$\astRep{\tKw{env}\tOp{()}}$)\jsOp{;}#
\end{lstlisting}

\caption{Builtin function.}

\label{get-impl}

\end{subfigure}

\begin{subfigure}{\columnwidth}
\lstset{language=JavaScript}
\begin{lstlisting}[
  resetmargins=true,
  linebackgroundcolor={%
  \ifnum\value{lstnumber}=2\color{gray!30}\fi
  \ifnum\value{lstnumber}=3\color{gray!30}\fi
  \ifnum\value{lstnumber}=4\color{gray!30}\fi
  \ifnum\value{lstnumber}=5\color{gray!30}\fi
  \ifnum\value{lstnumber}=6\color{gray!30}\fi
  \ifnum\value{lstnumber}=7\color{gray!30}\fi
  \ifnum\value{lstnumber}=8\color{gray!30}\fi
  \ifnum\value{lstnumber}=10\color{gray!30}\fi
  \ifnum\value{lstnumber}=11\color{gray!30}\fi
  \ifnum\value{lstnumber}=14\color{gray!30}\fi
  \ifnum\value{lstnumber}=15\color{gray!30}\fi
  \ifnum\value{lstnumber}=16\color{gray!30}\fi
  \ifnum\value{lstnumber}=17\color{gray!30}\fi
  \ifnum\value{lstnumber}=19\color{gray!30}\fi
  \ifnum\value{lstnumber}=20\color{gray!30}\fi
  }]
#\jsKw{let}~F~\jsOp{=}~\jsKw{function}\jsOp{(}resp\jsOp{)}~\jsOp{\{}#
  #\rtsFun{label}{\tRetlabel}#;
  #\rtsFun{enterSeq}{3}\jsOp{;}#
  #\rtsFun{let}{\astRep{\texttt{env}},\rtsFun{popArg}{}}\jsOp{;}#
  #\rtsFun{seqNext}{}\jsOp{;}#
  #\rtsFun{let}{\astRep{\texttt{resp}},\rtsFun{popArg}{}}\jsOp{;}#
  #\rtsFun{seqNext}{}\jsOp{;}#
  #\rtsFun{set}{\astRep{\texttt{out}},\astRep{\texttt{resp}}}\jsOp{;}#
  #out~\jsOp{=}~resp\jsOp{;}#
  #\rtsFun{pop}{}\jsOp{;}#
  #\rtsFun{pop}{}\jsOp{;}\tikzmark{ex-handler-return-cb}#
#\jsOp{\};}#

#\rtsFun{pushArg}{\astRep{\texttt{F}}}\jsOp{;}#
#\rtsFun{pushArg}{\astRep{\tOp{\texttt{'example.com'}}}}\jsOp{;}#
#\rtsFun{pushArg}{\astRep{\texttt{get}}}\jsOp{;}#
#\rtsFun{named}{\astRep{\texttt{r}}}\jsOp{;}#
#\jsKw{let}~r~\jsOp{=}~\jsKw{get}\jsOp{(\texttt{'example.com'},}~F\jsOp{)}\jsOp{;}\label{ex-handler-get-call}\tikzmark{ex-handler-get-call}#
#\rtsFun{pop}{}\jsOp{;}#
#\rtsFun{saveHandler}{0}\jsOp{;}\tikzmark{ex-handler-eof}#
\end{lstlisting}

\caption{Program.}
\label{get-example-program}

\end{subfigure}
\end{minipage}
\vrule\,\,
\begin{minipage}{0.54\columnwidth}
\begin{subfigure}{\columnwidth}
\footnotesize
\(
\def\arraystretch{1.2}
\begin{array}{@{}l}
\textrm{Tracing state} \\
\hline
\tikzmark{ex-handler-get-call-cfg}c = \tUnk  \\
\alpha = [F, \tOp{'example.com'}, \mathit{get}] \\
\cont = \contcase{Named}(r,\contcase{seq}([\tKw{let}~F\tOp{=}\tKw{env}\tOp{();}], [], \cdot)) \\
T = [ 0 \mapsto \tHandler{\_}{\_}{\tUnk} ] \\
\hline
\tikzmark{ex-handler-reg-cfg}c = \tKw{event}\tOp{('}\tKw{get}\tOp{','}\tKw{example.com/}\tOp{',1)} \\
\alpha = [] \\
\cont = \contcase{Named}(r,\contcase{seq}([\tKw{let}~F\tOp{=}\tKw{env}\tOp{();}], [], \cdot)) \\
T = [
0 \mapsto \cdots,
1 \mapsto \tKw{handler}\tOp{(}\tKw{env}\tOp{:}\tKw{env}\tOp{(),}\tKw{arg}: x,\tKw{body}\tOp{:}\tUnk\tOp{)}
] \\
\hline
\tikzmark{ex-handler-eof-cfg}c = \tKw{let}~F\tOp{=}\tKw{env}\tOp{();}\tKw{let}~r\tOp{=}\tKw{event}\tOp{('}\tKw{get}\tOp{','}\tKw{ex...}\tOp{',1);} \\
\alpha = [] \\
\cont = \cdot \\
T = [ 0 \mapsto \tHandler{\_}{\_}{c}, 1 \mapsto \cdots ]  \\
\hline
\tikzmark{ex-handler-call-cb-cfg}c = \tUnk \\
\alpha = [x, \tKw{env()}] \\
\cont = \cdot \\
T = [ 0 \mapsto \cdots, 1 \mapsto \cdots ] \\
\hline
\tikzmark{ex-handler-return-cb-cfg}c = \tRetlabel\tOp{:}~\tOp{\{}\tKw{let}~\mathit{env}\tKw{=}\tKw{env}\tOp{();}\tKw{let}~\mathit{resp}\tKw{=}x\tOp{();}\mathit{out}\tOp{=}\mathit{resp}\tOp{;\}} \\
\alpha = [] \\
\cont = \cdot \\
T = [ 1 \mapsto \tKw{handler}\tOp{(}\tKw{env}\tOp{:}\tKw{env}{()}\tOp{,}\tKw{arg}: x,\tKw{body}\tOp{:}\tUnk\tOp{)} ] \\
\hline
\tikzmark{ex-handler-save-cfg}c = \tRetlabel\tOp{:}~\tOp{\{}\tKw{let}~\mathit{env}\tKw{=}\tKw{env}\tOp{();}\tKw{let}~\mathit{resp}\tKw{=}x\tOp{;}\mathit{out}\tOp{=}\mathit{resp}\tOp{;\}} \\
\alpha = [] \\
\cont = \cdot \\
T = [ 0 \mapsto \cdots, 1 \mapsto \tHandler{\tKw{env}{()}}{x}{c} ]
\end{array}
\)
\caption{Execution trace.}
\label{get-example-trace}

\end{subfigure}
\end{minipage}

\begin{tikzpicture}[remember picture,overlay]
  \draw[dashed,-latex](pic cs:ex-handler-get-call) to (pic cs:ex-handler-get-call-cfg);
  \draw[dashed,-latex](pic cs:ex-handler-reg) to (pic cs:ex-handler-reg-cfg);
  \draw[dashed,-latex](pic cs:ex-handler-eof) to [bend right=30] (pic cs:ex-handler-eof-cfg);
  \draw[dashed,-latex](pic cs:ex-handler-call-cb) to (pic cs:ex-handler-call-cb-cfg);
  \draw[dashed,-latex](pic cs:ex-handler-return-cb) to (pic cs:ex-handler-return-cb-cfg);
  \draw[dashed,-latex](pic cs:ex-handler-save) to (pic cs:ex-handler-save-cfg);

\end{tikzpicture}

\caption{Event handler example. A simplified implementation of the \lstinline|get| function in
\sysname{}. The highlighted lines actually issue the request, and the other lines are needed for tracing.}
\end{figure}
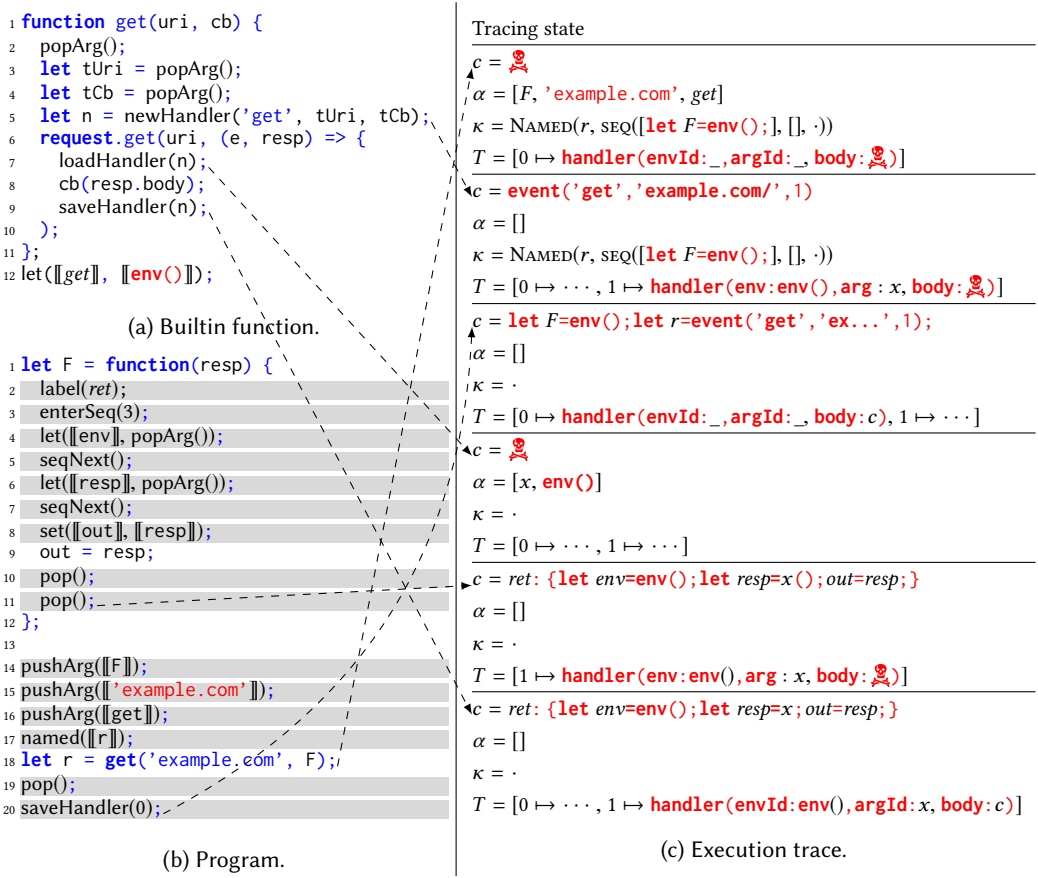

\subsection{Tracing Event Handlers}
\label{event-handlers}

\sysname{} provides programmers with an API of callback-based I/O functions.
Each function uses the runtime system to create an event handler
(\tKw{handler}) and issue an asynchronous event (\tKw{event}). The key
challenges are to 1)~manage multiple trace trees for multiple event handlers,
and 2)~support nested event handlers that capture non-local variables.

For example, the function \lstinline|get(url, callback)| issues an asynchronous
GET request to \lstinline|url| and calls the \lstinline|callback| function with
the response. To actually issue the HTTP request, \lstinline|get| uses a
function from a widely-used Node library called \lstinline|request.get|
(line~\ref{line:ex-actual-get}) (we elide error handling). To manage tracing,
\lstinline|get| relies on three helper functions that we add to the runtime
system (\cref{tracing-rts-events}):

\begin{enumerate}

\item We call function \rtsFunName{newHandler} immediately before
registering an event handler in JavaScript. This helper function reflects the
newly created event handler by 1)~creating a new \tKw{handler} and 2)~setting
the current trace ($c$) to an \tKw{event} expression. Note that the body of the
\tKw{handler} is initialized to \tUnk{}. However, as long as the event triggers
as response, that \tUnk{} will be replaced with the trace of the event handler.

\item We call the function \rtsFunName{loadHandler} immediately after receiving
an event. This function prepares the runtime to trace the callback by 1)~pushing
the traces of its environment and argument onto the argument stack, and
2)~setting the current trace ($c$) to the trace in the handler ($\tKw{body}$).

\item Finally, we call the function \rtsFunName{saveHandler} after the callback
returns to store the current trace back into the handler. Therefore, if the
callback executes multiple times, the trace in the handler will be restored and
grown in each call.

\end{enumerate}
The pattern that \lstinline|get| employs applies to all other callback
functions.

\Cref{get-example-program} is an example program that makes a request
using \lstinline|get|. As in prior examples, the figure shows the output
of the compiler. \Cref{get-example-trace} shows the state of the
tracing runtime at the program executes, including the set of event handlers.

\section{Compiling Traces to Rust}
\label{traces-to-rust}

We now present how we compile traces to Rust, which has two major steps: 1)~We
impose CPU and memory limits on the program, and 2)~We address the mismatch
between the types of values in traces (which is dynamically typed) and Rust
(which is statically typed). To address the latter, we inject all values into a
\emph{dynamic type}~\cite{dynamic-type} and use \emph{arena allocation} to
simplify reasoning about Rust's lifetimes. An arena---by design---can only free
all allocated values at once. Our runtime system exploits the fact that
serverless functions have transient memory and simply clears the arena after
each request to the serverless function.

\begin{figure}

\lstset{language=rust}
\begin{lstlisting}
#[derive(Copy, Clone)]
pub enum Dyn<'a> {
  Int(i32),
  Bool(bool),
  Undefined,
  Object(&'a RefCell<Vec<'a, (&'a str, Dyn<'a>)>>),@\label{refcell-ex}@
}

impl<'a> Dyn<'a> {
  pub fn add(&self, other: &Dyn<a'>) -> Dyn<'a> {
    match (self, other) {
      (Dyn::Int(x), Dyn::Int(y)) => Dyn::Int(x + y),
      ...
    }
  }
}
\end{lstlisting}

\caption{A fragment of the dynamic type that \sysname{} uses to
represent trace values.}

\label{type-dynamic}

\end{figure}

\subsection{Static Types and Arena Allocation}

Compiling the dynamically typed trace program to statically-typed Rust presents three
separate issues.

\paragraph{Dynamic type}

In JavaScript, we can write expressions such as \lstinline|1 + true| (which
evaluates to $2$). However, that program produces a type error in Rust. To
address this problem, we use the well-known technique of defining a
\emph{dynamic type} for trace values, which enumerates all
possible types that a value may have. \Cref{type-dynamic} shows the Rust code
for a simplified fragment of the dynamic type that we employ. The cases of this
dynamic type includes simple values, such as numbers and booleans, as well as
containers such as objects. In addition, the dynamic type implements methods
for all possible operations for all cases in its enumeration, and these methods
may fail at runtime if there is a genuine type error.
Therefore, we would compile \lstinline|1 + true| to the following Rust code:
\begin{lstlisting}
Dyn::Int(1).add(Dyn::Bool(true))
\end{lstlisting}
The \lstinline|add| method implements the type conversions
 necessary for JavaScript.

\paragraph{Aliased, mutable pointers}

The Rust type system guarantees that all mutable pointers are unique, or
\emph{own} the values that they point to. Therefore, it is impossible for two
mutable variables to point to the same value in memory. However, JavaScript
(and other dynamic languages) have no such restrictions, and neither does the
trace language. Rust's restriction allows the language to ensure that concurrent programs
are data race free. However, for code that truly requires multiple mutable
references to the same object, the Rust standard library has a container type
(\lstinline|RefCell|) that dynamically checks Rust's ownership rules, but
prevents the value from crossing threads.
Since trace programs execute in a single-threaded manner, we can use
\lstinline|RefCell| to allow aliases. For example, the dynamic type represents
objects as a vector of key-value pairs stored inside a \lstinline|RefCell|
(\cref{type-dynamic}, line~\ref{refcell-ex}).

\paragraph{Lifetimes and arena allocation}

Variables in Rust have a statically-determinate lifetime, and the value stored
in a variable is automatically deallocated once the lifetime goes out of scope.
In contrast, variables in a trace tree may be captured in environment objects,
and thus have a lifetime that is not statically known. There are a variety of
workaround in Rust, e.g., reference counting and dynamic borrow checking.
However, the Rust type system does not guarantee that programs that use these
library features do not leak memory (e.g., due to reference cycles). Therefore,
reference counting is not safe to use in \sysname{}.

To solve this, \sysname{} uses an \emph{arena} to store the values of a running
trace program. Arena allocation simplifies lifetimes, since the lifetime of all
values is the lifetime of the arena itself. This is why our dynamic type has
single lifetime parameter (\lstinline|'a| in \cref{type-dynamic}), which is the
lifetime of the arena in which the value is allocated. Another benefit of
arenas is that they support very fast allocation. However, it is not possible
to free individual values in an arena. Instead, the only way to free a value in
an arena is to free all values in the arena.

Fortunately, the serverless execution model gives us a natural point to
allocate and clear the arena. \sysname{} allocates an arena for each request
and clears it immediately after the function produces a response. This is safe
to do because serverless functions must tolerate transient memory.

\subsection{Bounding Memory and Execution Time}

Serverless computing relies on bounding the CPU and memory utilization of
serverless functions. The arena allocator makes it easy to impose a memory
bound: all values have the same lifetime as the allocator, and we impose a
maximum limit on the size of the arena. Imposing a CPU utilization limit is more
subtle, since \sysname{} can run several trace programs in the same process,
thus we cannot accurately account for the CPU utilization for an individual
request. Instead, the trace-to-Rust compiler uses an instruction counter, which
it increments at the top of every loop and at the end of every invocation of the
state machine, and we bound the number of Rust statements executed.

\section{The \sysname{} Invoker}
\label{rts}

The \sysname{} invoker can process an event in one of two ways. 1)~The invoker
manages a pool of containers that run the serverless function, and it can
dispatch an event to an idle container, start a new container (up to a
configurable limit), and stop idle containers. 2)~The invoker can also dispatch
events to a compiled trace tree, which bypasses the container. Which method the
invoker uses depends on it being within one of two possible modes. 1)~In
\emph{tracing mode}, the invoker does not have a compiled trace tree and thus
processes all events using containers. It configures the first container it
starts for the function to build a trace tree, and after a number of events, it
compiles the trace to Rust. 2)~In \emph{containerless mode}, the invoker
dispatches events to the compiled trace tree. Ideally, the invoker stays in
containerless mode indefinitely, but it is possible for the invoker to receive
an event that leads to unknown behavior (\tUnk). When this occurs, it reverts
back to tracing mode, and sends the event that triggered \tUnk{} to a container.
To avoid ``bouncing'' between containerless and tracing modes, the invoker
keeps count of how many times it has bounced, and eventually enters
\emph{container mode}, where it ceases tracing and behaves like an ordinary
invoker.

\section{Evaluation}
\label{evaluation}

Our primary goal is to determine if \sysname{}
can reduce the latency and resource usage of typical serverless functions.

\paragraph{Benchmark Summary}

We develop \dataNumBenchmarks{} benchmarks:

\begin{enumerate}

    \item \emph{authorize}: a serverless function is equivalent to the running
    example in the paper (\cref{login-example-js}). It receives as input a
    username and password, fetches the password database (represented as a JSON
    object), and validates the input.

    \item \emph{upload}: a serverless function that uploads a file to cloud
    storage. It receives the file in the body of a POST request and issues a POST
    request to upload it.

    \item \emph{status}: a serverless function that updates build status
    information on GitHub. i.e., it can add a \ding{51} or \ding{55} next to a
    commit, with a link to a CI tool. The function takes care of mapping a
    simple input to the JSON format that the GitHub API requires.

    \item \emph{banking}: a serverless function that simulates a banking
    application, with support for deposits and withdrawals (received over POST
    requests). It uses the Google Cloud Datastore API with transactional
    updates.

    \item \emph{autocomplete}: a serverless function that implements autocomplete.
    Given a word as input, it returns a number of completions.

    \item \emph{maze}: a relatively computationally expensive serverless
    function, that finds the shortest path between two points in a maze on each
    request.

\end{enumerate}

\paragraph{Experimental Setup}

We run the \sysname{} invoker on a six-core Intel Xeon E5-1650 with 64 GB RAM.
We send events from an identical machine on the same rack, connected to the
invoker via a 1 GB/s connection. Serverless platforms impose memory and CPU
limits on containers. We allocate \dataContainerCPULimit{} and
\dataContainerMemoryLimit{} to each container.

A number of our benchmarks rely on external services (e.g., Github and Google
Cloud Datastore). We tested that they actually work. But, in the experiments
below, we send requests to a mock server. The experiments stress \sysname{{}
and issue thousands of requests per second, and our API keys would be
rate-limited or even blocked if we used the actual services.

\begin{figure*}

\begin{subfigure}{0.32\textwidth}
\begin{tikzpicture}
\node{\pgfimage[width=\columnwidth]{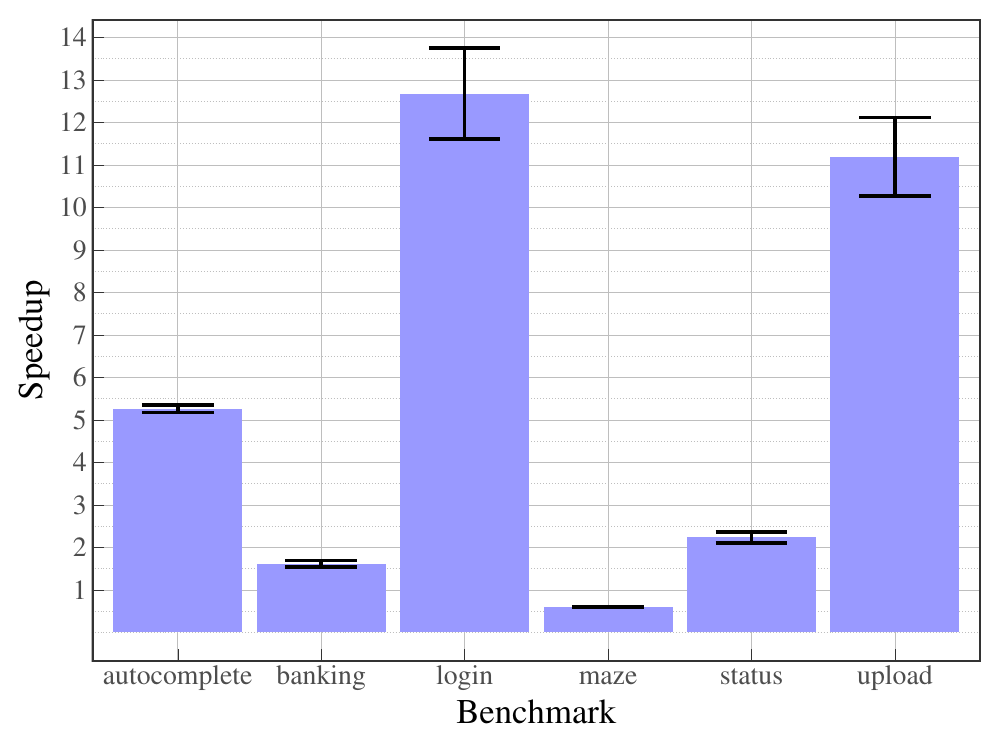}};
\end{tikzpicture}
\caption{Speedup of \sysname{}.}
\label{warm-processing-time}
\end{subfigure}
\begin{subfigure}{0.32\textwidth}
\begin{tikzpicture}
\node{\pgfimage[width=\columnwidth]{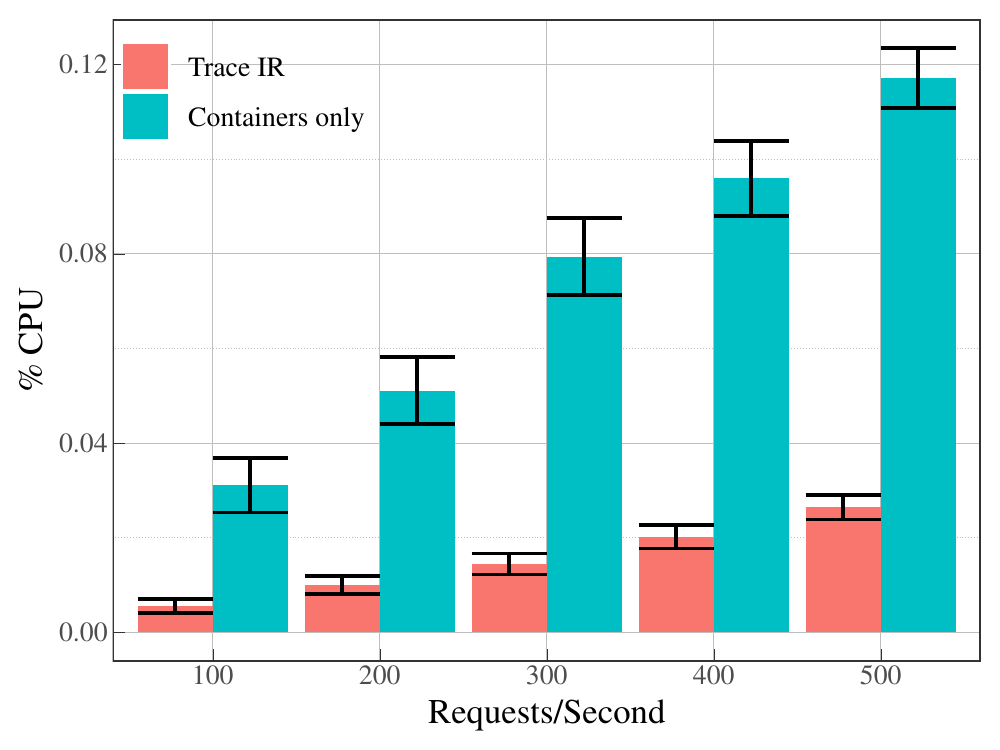}};
\end{tikzpicture}
\caption{CPU utilization of \emph{authorize}.}
\label{cpu-utilization}
\end{subfigure}
\begin{subfigure}{0.32\textwidth}
\begin{tikzpicture}
\node{\pgfimage[width=\columnwidth]{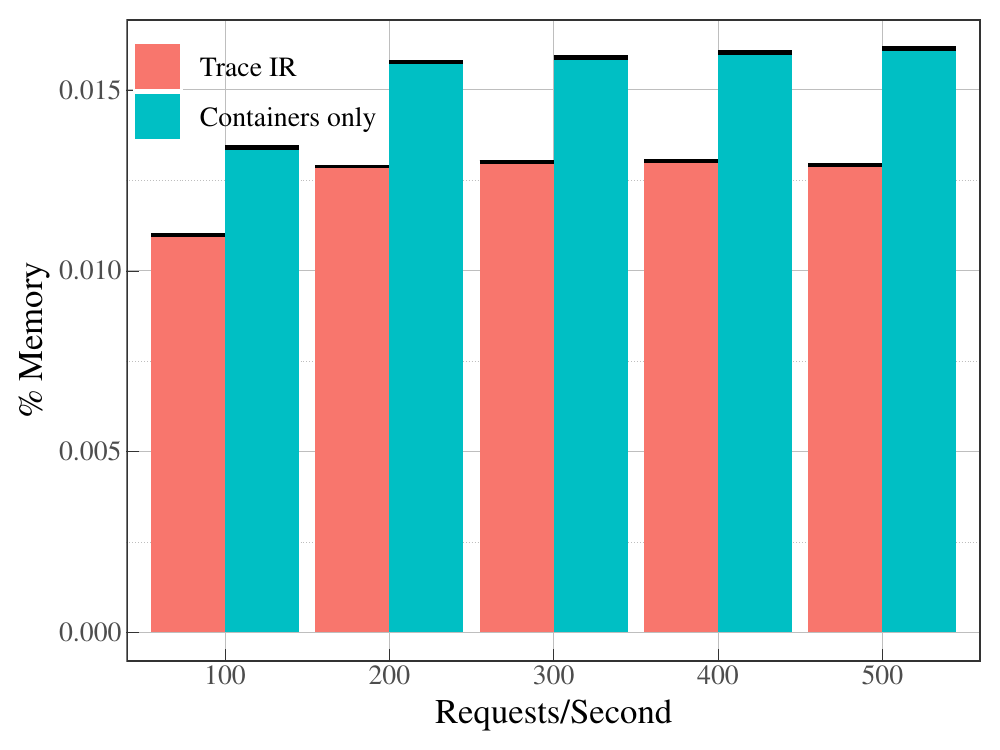}};
\end{tikzpicture}
\caption{Memory utilization of \emph{authorize}.}
\label{mem-utilization}
\end{subfigure}
\caption{Speedup, CPU utilization, and memory utilization. The error bars show
the 95\% confidence interval.}

\end{figure*}

\subsection{Steady-State Performance}
\label{steady-state}

For our first experiment, we measure invoker performance with and without
\sysname{}. We send events using ten concurrent event streams, where each
stream immediately issues another event the moment it receives a response. We
measure end-to-end event processing latency and report the speedup with
\sysname{}.

We run each benchmark for \dataWarmBenchmarkRunningTime{} seconds and we start
measurements after 30 seconds. This gives \sysname{} time to extract the trace
program, run the trace-to-Rust compiler, and start handling all events in Rust.
When running without \sysname{}, the experiments ensure that the event arrival
rate is high enough that containers are never idle, thus are never
stopped by the invoker. In addition, the invoker does not pause containers,
which adversely affects latency~\cite{shahrad:micro-faas}.
\Cref{warm-processing-time} shows the mean speedup for each benchmark with
\sysname{}. In five of the six benchmarks, \sysname{} is significantly faster,
with speedups ranging from \dataMinActualSpeedup{} to \dataMaxSpeedup.

The outlier is the \emph{maze} benchmark, which runs 60\% slower with
\sysname{}. \emph{Maze} is much more computationally expensive than the other
benchmarks. It also doesn't perform any I/O, although \emph{autocomplete} does
not either. With some engineering, it should be possible to make \emph{maze}
run faster. We believe that the reason for the slowdown is that \emph{maze}
uses a JavaScript array as a queue. JavaScript JITs support multiple array
representations and optimize for this kind of behavior. However, the
implementation of dequeueing (the \lstinline|.shift| method) in our Rust
runtime system is an $O(n)$ operation. We could improve our performance on
\emph{maze}, but there will always be certain functions---particularly those
that are compute-bound---where a JavaScript JIT outperforms the \sysname{}
approach. One approach that the invoker could use is to actively measure
performance, and if it finds that the Rust code is performing worse,
revert to containerization permanently on that function. However,
the performance characteristics of \emph{maze} is more subtle, as the
next experiment shows.

\begin{figure*}
\begin{subfigure}[t]{0.49\textwidth}
\begin{tikzpicture}
\node{\pgfimage[width=0.97\columnwidth]{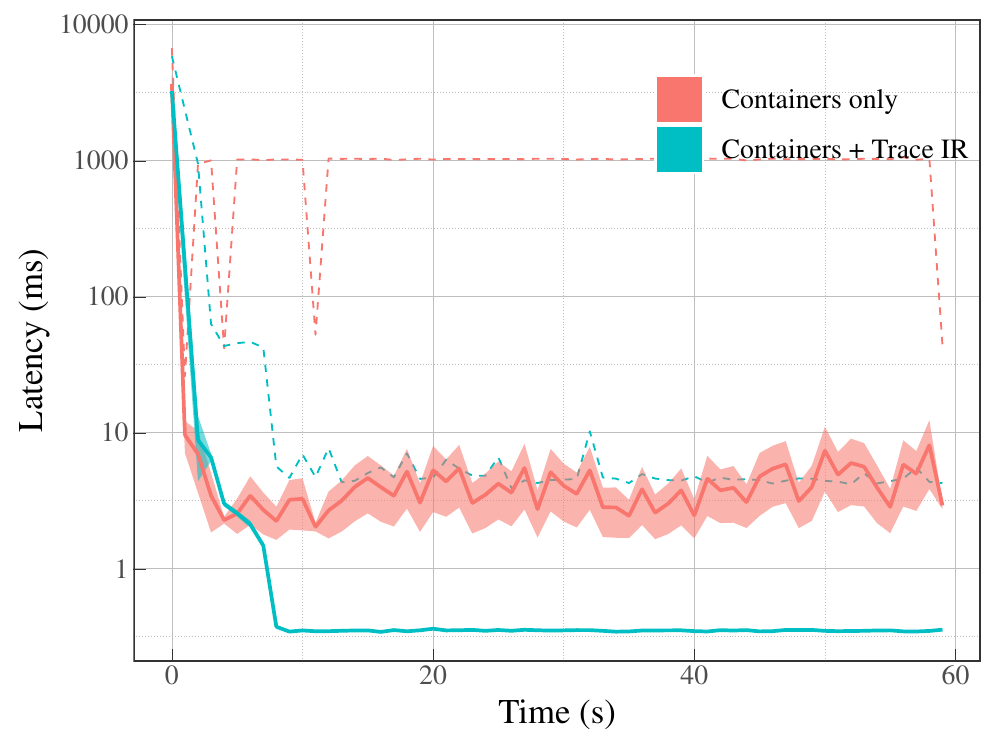}};
\end{tikzpicture}
\caption{The \emph{upload} benchmark.}
\label{upload-processing-time}
\end{subfigure}
\begin{subfigure}[t]{0.49\textwidth}
\begin{tikzpicture}
\node{\pgfimage[width=0.97\columnwidth]{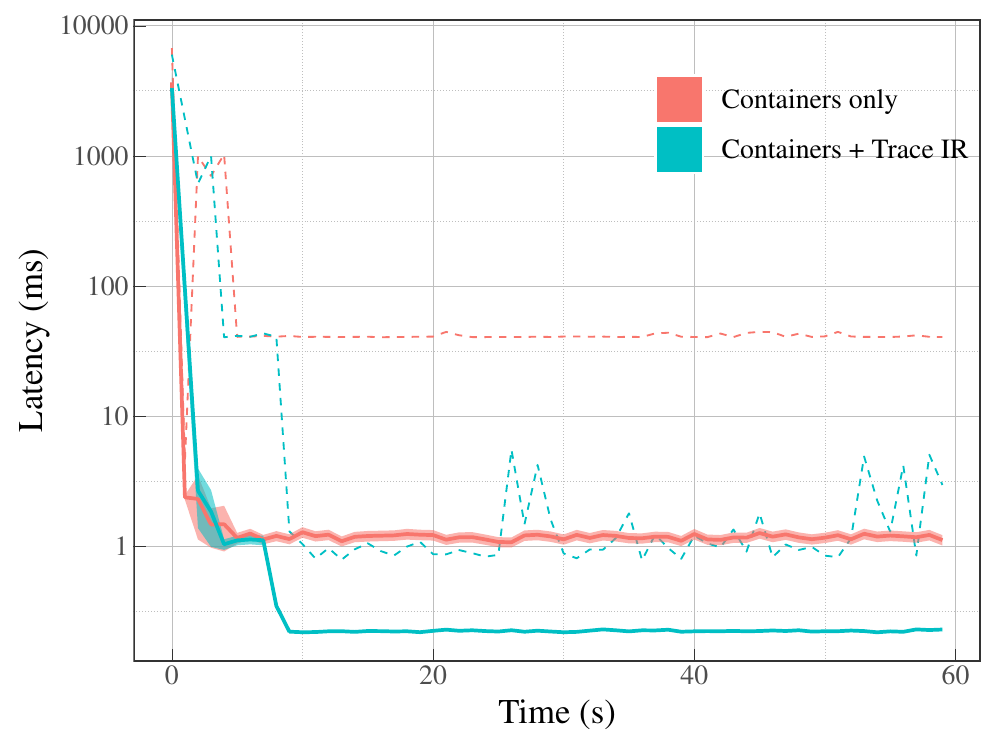}};
\end{tikzpicture}
\caption{The \emph{autocomplete} benchmark.}
\label{autocomplete-processing-time}
\end{subfigure}

\begin{subfigure}[t]{0.49\textwidth}
\begin{tikzpicture}
\node{\pgfimage[width=0.97\columnwidth]{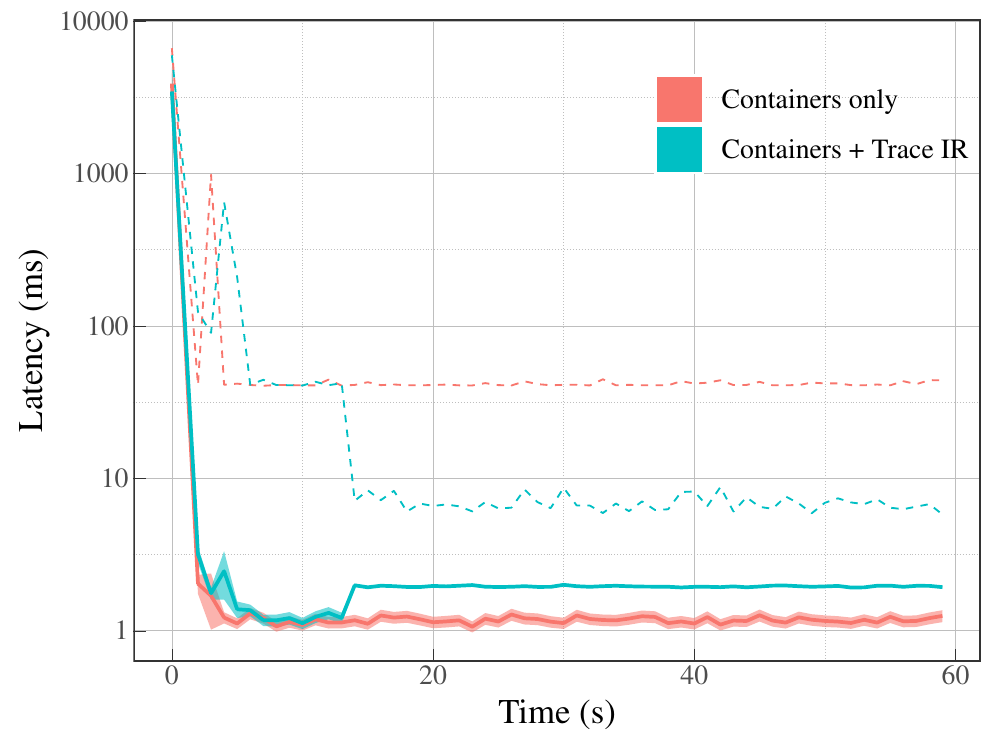}};
\end{tikzpicture}
\caption{The \emph{maze} benchmark.}
\label{maze-processing-time}
\end{subfigure}
\begin{subfigure}[t]{0.49\textwidth}
\begin{tikzpicture}
\node{\pgfimage[width=0.97\columnwidth]{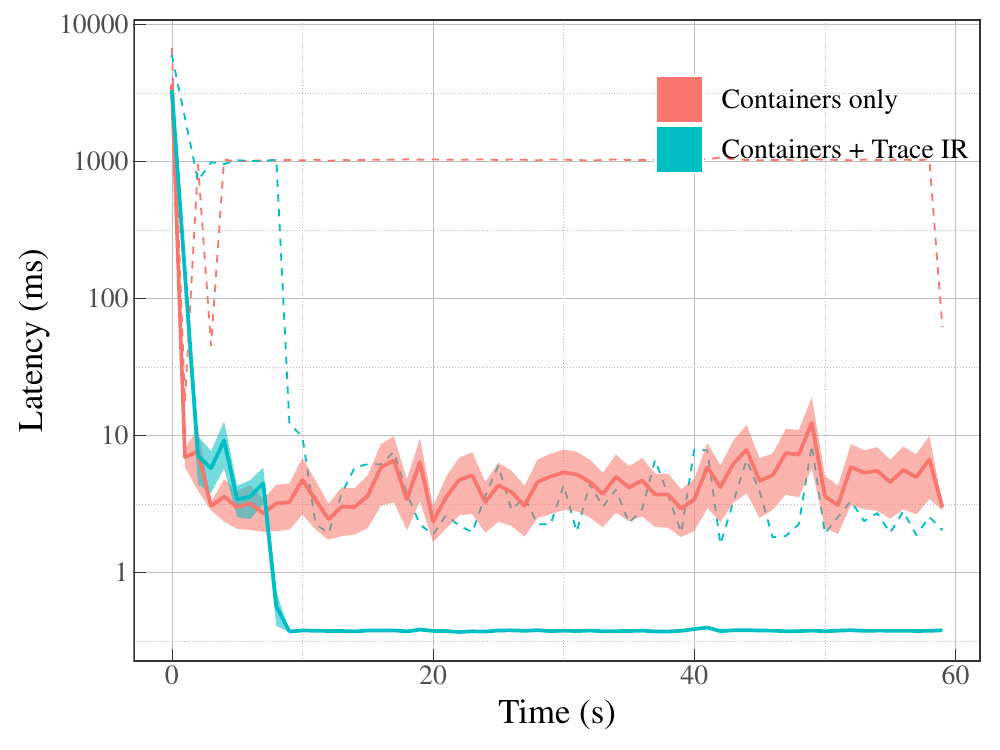}};
\end{tikzpicture}
\caption{The \emph{authorize} benchmark.}
\label{authorize-processing-time}
\end{subfigure}

\begin{subfigure}[t]{0.49\textwidth}
\begin{tikzpicture}
\node{\pgfimage[width=0.97\columnwidth]{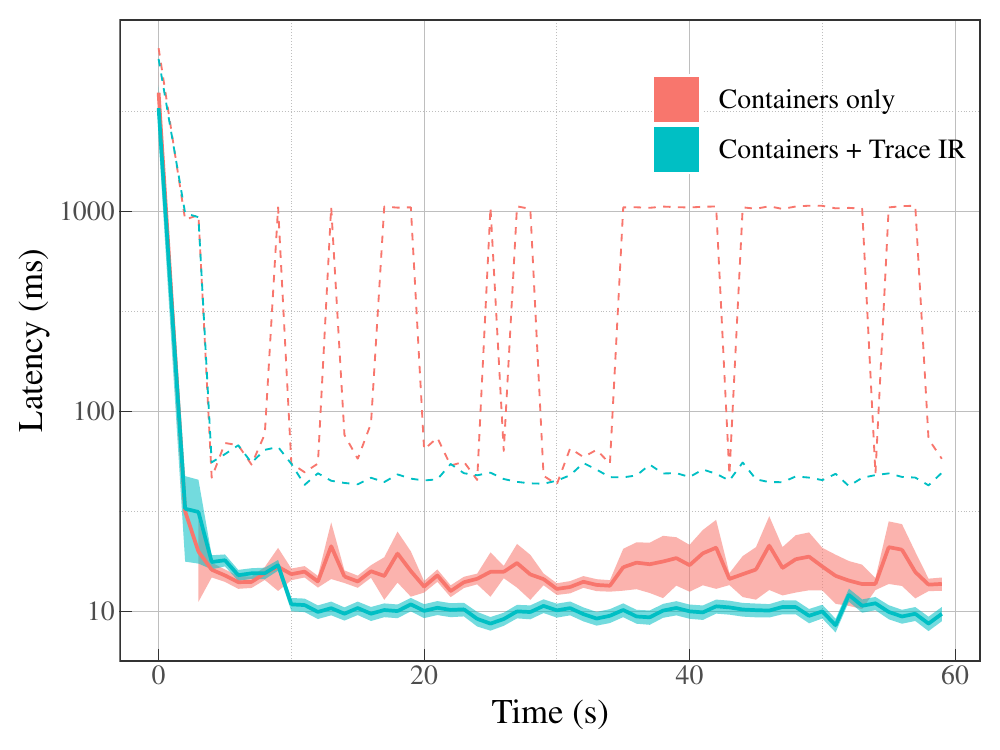}};
\end{tikzpicture}
\caption{The \emph{banking} benchmark.}
\label{banking-processing-time}
\end{subfigure}
\begin{subfigure}[t]{0.49\textwidth}
\begin{tikzpicture}
\node{\pgfimage[width=0.97\columnwidth]{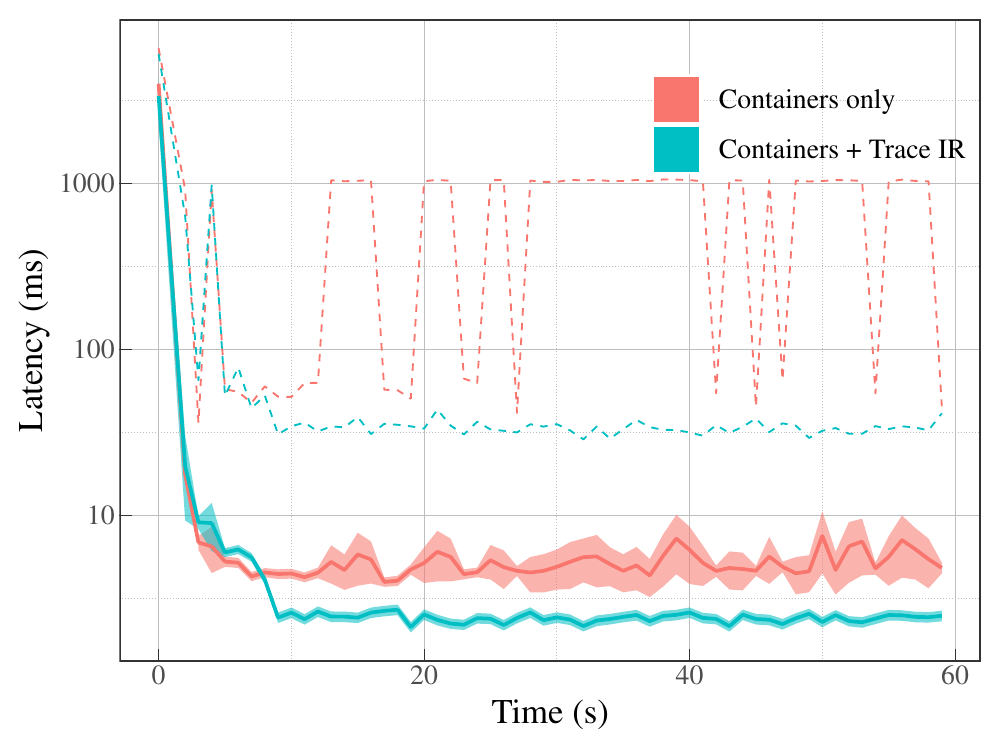}};
\end{tikzpicture}
\caption{The \emph{status} benchmark.}
\label{status-processing-time}
\end{subfigure}

\caption{Cold-to-warm performance with and without \sysname{}. Each experiment
runs for one minute and begins with no containers loaded. Each graph summarizes
the latency of events issued at a point in time, with $t = 0$ is the start of
the experiment. The solid lines show the mean event latency, with the 95\%
confidence interval depicted by the shaded region around the mean. The dotted
lines show the maximum latency.}

\label{cold-to-warm-graphs}

\end{figure*}

\subsection{Cold-to-Warm Performance}
\label{cold-to-warm}

Our second set of experiments examine the behavior of \sysname{} under cold
starts. As in the previous section, we run each benchmark with and without
\sysname{}, issuing events using ten concurrent event streams. We run each
experiment for one minute, starting with no running containers.
\Cref{cold-to-warm-graphs} plots the mean and maximum event processing latency
over time.

Let us examine \emph{upload} in detail (\cref{upload-processing-time}):
\begin{itemize}

  \item \textbf{Cold starts:} At $t=0$, \sysname{} and container-only both
  exhibit cold starts (very high latency) as the containers warm up. \emph{Note
  that the latency ($y$-axis) is on a log scale}.

  \item \textbf{Warm starts:} Since there are ten concurrent event streams, both
  cases start up the maximum number of containers (six), where one of the containers
  runs tracing for \sysname{}. Once they are all started, mean latency for
  both invokers dips to about 5 ms. However, tracing does incur some overhead,
  and we can see that the mean latency for \sysname{} takes slightly longer to
  reach 5 ms.

  \item \textbf{\sysname{} starts:} However, in the \sysname{} case, within eight
  seconds, the tracing container receives enough events for \sysname{} to
  extract the trace, compile it, and start processing events in Rust. Thus
  the mean latency for \sysname{} \emph{dips again} to 0.3 ms after eight
  seconds.

  \item \textbf{Variability:} The plot also shows the event processing time has
  higher variability with containers. This occurs because there are ten
  concurrent connections and only six containers (one for each core) thus some
  events have to be queued. \sysname{} runs in a single process, with one
  physical thread for each core. However, the Rust runtime system (Tokio)
  supports non-blocking I/O and is able to multiplex several running trace
  programs on a single physical thread, thus can process more events
  concurrently.

\end{itemize}
The plots for the other benchmarks, with the exception of \emph{maze}, also
exhibit this ``double dip'' behavior: first for warm starts, and then again
once \sysname{} starts its language-based sandbox.

As discussed in \cref{steady-state}, \emph{maze} is relatively
compute-intensive, and \sysname{} makes its mean latency worse (when $t>8$ in
\cref{maze-processing-time}). However, at the same time, the maximum latency
(dashed green line) is significantly lower with \sysname{} than without! Since
\emph{maze} does not perform any asynchronous I/O, we cannot attribute this
behavior to nonblocking I/O. It is hard to pinpoint the root cause of this
behavior. One possibility is the difference is memory management: within
the container, the program runs in a JavaScript VM that incurs brief GC pauses,
whereas \sysname{} uses arena allocation, and clears the arena immediately
after each response. However, this is a conjecture, and there are several
differences between \sysname{} and container-only execution.

\subsection{Resource Utilization}

Our third experiment examines CPU and memory utilization. We use the
\emph{authorize} benchmark and vary the number of requests per second. The
maximum number of requests per second that we issue is 500, because a higher
request rate exceeds the rate at which containers can service requests. We
examine resource utilization after the cold start period. As shown in
\cref{cpu-utilization}, \sysname{} has a lower CPU utilization than containers
by a factor of \dataHeadlineCPU. \cref{mem-utilization} shows that
\sysname{} lowers memory utilization by a factor of \dataHeadlineMem.

\subsection{An Alternative to Cold Starts}

\sysname{} does not eliminate cold start latency, since it needs the function
to run in a container to build the trace program. However, traced programs present
a new opportunity: since they are more lightweight than containers, the
invoker can keep them resident significantly longer. For example, on our
experimental server, running \emph{authorize} in $100$ containers consumes 1.6
GB of physical memory. In contrast, an executable that contains $100$ copies of
the trace produced by \emph{authorize} is $10$ MB. In \sysname{}, the
arena allocator frees memory after a response, thus the only memory consumed by
a function that is loaded and idle, is the memory needed for its code, and for
its entry in a dispatch table, which maps a URL to a function pointer.

At scale, a single invoker would not be able to have trace programs loaded for
\emph{all} serverless functions. Moreover, a platform running several
\sysname{} invokers could benefit from a mechanism that allows a trace program
built on one node to be shared with other nodes. We leave this for future work.

\section{Discussion}
\label{discussion}

The design of \sysname{} raises several questions, which we discuss below.

\paragraph{Security}

The design of \sysname{} is motivated by the desire to minimize the size of its
trusted computing base (TCB). The only trusted component in \sysname{} is the
invoker (\cref{rts}), which is a relatively simple system. The most
sophisticated parts of \sysname{} are untrusted: 1)~the tracing infrastructure
(\cref{tracing}) runs within an untrusted container, and can be compromised
without affecting the serverless platform; 2)~the trace-to-Rust compiler
(\cref{traces-to-rust}) may have a bug that produces unsafe code, but such a
bug would either be caught by Rust or by simple extra verification in the
invoker (loops must increment the instruction counter, and the function cannot
load arbitrary libraries). We do place trust in large piece of third-party
code: the Rust compiler and runtime system. However, we argue that Rust
is increasingly trusted by other security-critical applications (e.g., Amazon
Firecracker).

\sysname{} allows running untrusted code from multiple parties in the same
address space, which means that Spectre attacks are a concern~\cite{spectre}.
However, we believe there are a few mitigating factors. First, the \sysname{}
runtime does not give the trace language direct access to timers. JavaScript programs that need a
timer are thus confined to containers. Second, \sysname{} limits how many
instructions a trace program can execute. Programs that need to run longer are
also confined to containers. We do not claim that our approach is immune to
side-channel attacks, but it may be possible to mitigate them by restricting
the resources available to programs in the language-based sandbox. \sysname{}
can also be combined with process-based isolation for better defense, similar
to Boucher et al.~\cite{microin}.

\paragraph{Alternative designs}

We can imagine other approaches to serverless function acceleration. For
example, we could run a JavaScript VM that runs out of the container with a
restricted API (similar to CloudFlare Workers), and fall back to the
containerized JavaScript VM if the serverless function performs an unsupported
operation. We could also compile a fragment of JavaScript directly to Rust, and
omit tracing entirely. The former approach would require trust in a larger
codebase, whereas the latter approach is likely to support fewer programs.

\paragraph{How much tracing is necessary?}

This paper does not address some important questions that affect the performance of
\sysname{}. For example, how many requests need to be traced to get a
program that is sufficiently complete? Our evaluation uses a fixed number for
simplicity. To do better, we need to develop a larger suite of serverless
functions. We conjecture that the answer will depend on the function, so an
adaptive strategy could be most effective.

\paragraph{Growing the API}

The \sysname{} API is small, but already usable. Our benchmark programs use
typical external services, such as the GitHub API and Google Cloud Datastore.
Growing the API with additional functions does require work, for each added
function requires: 1)~The function has to be reimplemented in Rust and 2)~the
JavaScript implementation of the function needs a tracing shim. It should be
possible to write a tool that automatically generates the tracing shim in
JavaScript, since they all follow the same recipe. However, the Rust
reimplementation needs to be carefully built to ensure safety and JavaScript
compatibility.

\section{Related Work}
\label{related}

\paragraph{Serverless computing performance}

Serverless computing and container-based platforms in general have high
variability in performance, and several systems have tried to address
performance problems in a variety of ways. SAND~\cite{istemi:sand} uses
process-level isolation to improve the performance of applications that compose
several serverless functions together; X-Containers~\cite{x-containers:shen}
develops a new container architecture to speed up arbitrary microservices;
MPSC~\cite{aske:mspc} brings serverless computing to the edge;
Costless~\cite{costless} helps programmers explore the tradeoff between performance
and cost; and GrandSLAm~\cite{slam} improves microservice throughput by dynamic
batching. The \sysname{} approach differs from these solutions because it uses
speculative acceleration techniques to bypass the container when possible. As
long as the application code can be analyzed for tracing, a system like
\sysname{} can complement the aforementioned approaches.

\sysname{} exploits the fact that many serverless platforms rely on the
programmer to ensure that their functions that are idempotent and tolerate
transient in-memory state~\cite{jangda:lambda-lambda,obetz:serverless-events}.
In contrast, platforms such as Ambrosia~\cite{goldstein:ambrosia} provide a
higher-level abstraction and relieves programmers from thinking about these
low-level properties.

Boucher et al.~\cite{microin} present a serverless platform that requires
programmers to use Rust. As we discussed in \cref{introduction}, Rust has a
steep learning curve and---more fundamentally---Rust does not guarantee resource
isolation, deadlock freedom, memory leak freedom, and other critical safety
properties~\cite{rust-unsafe}. \sysname{} allows programmers to continue using
JavaScript and compiles their code to Rust. Moreover, the compiler ensures that
the output Rust code does not have deadlocks, memory leaks, and so on.

\paragraph{Tracing and JITs}

\sysname{} compiles dynamically generated execution trace trees, which is an
idea with a long history. Bulldog~\cite{ellis:bulldog} is a compiler that
generates execution traces statically, and uses these longer traces to produce
better code for a VLIW processor. TraceMonkey~\cite{tracemonkey} is a tracing
JIT for JavaScript that works with \emph{intra}procedural execution traces. It
was introduced in Firefox 3.5, but removed in Firefox 11. Spur~\cite{spur} is
an interprocedural tracing JIT for the Microsoft Common Intermediate Language
(CIL), thus it can generate traces that cross source-language boundaries.
RPython~\cite{bolz:metatracing} is a meta-tracing JIT, that allows one to write
an annotated interpreter, which RPython turns into a tracing JIT. In contrast,
Truffle~\cite{truffle} partially evaluates an interpreter instead of
meta-tracing. Tracing in \sysname{} differs from prior work in two key ways.
1)~Since the target language is a high-level language (Rust), the language of
traces is high-level itself. 2)~\sysname{} is designed for serverless
execution, and naively restarts the serverless function in a container when it
goes off trace, whereas prior work has to seamlessly switch between
JIT-generated code and the interpreter.

\paragraph{Operating systems}

There are a handful of research operating systems that employ language-based
sandboxing techniques to isolate untrusted code from a trusted kernel.
Processes in Singularity~\cite{singularity-sealed} are written in managed languages and disallow
dynamically loading code. SPIN~\cite{spin} and
VINO~\cite{seltzer:vino} allows programs to dynamically extend the kernel with
extensions that are checked for safety. Our trace language is analogous to an extension
written in a safe language. However, we do not ask
programmers to write traces themselves. Instead, we generate traces from executions
within a container. Moreover, \sysname{} switches between language-based and
container-based sandboxing as needed.

\paragraph{Other domain-specific accelerators}

There are other accelerators that translate programs to an intermediate representation. Weld~\cite{weld}
generates and optimizes IR code from data analytics applications that mix
several libraries and languages, and Numba~\cite{numba} accelerates Python and
NumPy code by JITing methods. Unlike \sysname{}, these systems do not employ
tracing. TorchScript~\cite{torchscript} is a trace-based accelerator for
PyTorch, though it places several restrictions on the form of Python code in a
model. All these accelerators, including \sysname{}, exploit domain-specific
properties to achieve their speedups. However, the domain-specific properties
of serverless computing are very different from data analytics, scientific
computation, and deep learning, thus \sysname{} uses serverless-specific
techniques that do not apply to these other domains.

\paragraph{Serverless as HPC}

There are a number of projects that use serverless computing for ``on-demand
HPC''~\cite{jonas:pywren,fouladi:excamera,fouladi:gg,ao:sprocket,
lee2018evaluation}. The current implementation of \sysname{} is unlikely to help
in these use-cases because many of them rely on native binaries. Moreover, the
code that we generate from trace programs is less efficient than a JavaScript JIT
on computationally expensive benchmarks. However, for short-running, I/O
intensive applications, our evaluation shows that \sysname{} can improve
performance significantly.

\section{Conclusion}
\label{conclusion}

This paper introduces the idea of \emph{language-based serverless function
acceleration}, which executes serverless functions in a language-based sandbox.
Our technique is speculative: all functions cannot be accelerated, but we can
detect acceleration failures at runtime, abort execution, and fallback to
containers. It is generally unsafe to naively restart arbitrary programs,
especially programs that interact with external services. However, our approach
relies on the fact that serverless functions must already be idempotent,
short-lived, and tolerate arbitrary restarts. Serverless platforms already
impose these requirements for fault tolerance, but we exploit these
requirements for acceleration.

We also present \sysname{}, which is a serverless function accelerator that
works by dynamically tracing serverless functions written in JavaScript. The
design of \sysname{} is driven by a desire to minimize the size of the TCB.
However, other accelerator designs are possible and may lead to different
tradeoffs.

\bibliography{main}

\end{document}